\documentclass[aps,prb,floatfix,amsmath,amssymb,preprint,eqsecnum,nofootinbib,superscriptaddress]{revtex4}
\usepackage{graphicx}
\usepackage{dcolumn}
\usepackage{bm}
\usepackage{epstopdf}
\usepackage{setspace}   

\newcommand{\beq}{\begin{equation}}
\newcommand{\eeq}{\end{equation}}
\newcommand{\ba}{\begin{array}{ccc}}
\newcommand{\ea}{\end{array}}
\newcommand{\nn}{\nonumber}
 \renewcommand{\d}{\partial}
\def\bea{\begin{eqnarray}}
\def\eea{\end{eqnarray}}

\def\prl{\parallel}

\def\<{\langle}
\def\>{\rangle}
\def\mx{\mathrm{x}}
\def\mm{\mathrm{m}}
\usepackage{amsmath}
\usepackage{amssymb}
\usepackage{color}
\begin{document}

\title{Entanglement Entropy of Systems with Spontaneously Broken Continuous Symmetry}

\author{Max A. Metlitski}
\affiliation{Kavli Institute for Theoretical Physics, University of California, Santa Barbara, CA 93106}

\author{Tarun Grover}
\affiliation{Kavli Institute for Theoretical Physics, University of California, Santa Barbara, CA 93106}

\date{\today\\
\vspace{1.6in}}
\begin{abstract}
We study entanglement properties of systems with spontaneously broken
continuous symmetry. We find that in addition to the expected area law
behavior, the entanglement entropy contains a subleading contribution
which diverges logarithmically with the subsystem size in agreement with
the Monte Carlo simulations of A. Kallin {\it et. al.}
(Phys.~Rev.~B {\bf 84}, 165134 (2011)). The coefficient of the logarithm is a universal
number given simply by $\frac{N_G \,(d-1)}{2}$ where $N_G$ is  the
number of Goldstone modes and $d$ is the spatial dimension.
This term is present even when the subsystem boundary is straight and
contains no corners, and its origin lies in the interplay of Goldstone
modes and restoration of symmetry in a finite volume. We also compute
the ``low-energy" part of the entanglement spectrum and show that it
has the same characteristic ``tower of states" form as the physical
low-energy spectrum obtained when a system with spontaneously broken
continuous symmetry is placed in a finite volume.
\end{abstract}

\maketitle

\section{Introduction}
In recent years there has been a lot of theoretical interest in entanglement properties of quantum states of matter. Entanglement has proved to be a useful probe of non-local correlations for both gapless and gapped systems. The most commonly used characterization of entanglement is the entanglement entropy $S = -\mathrm{tr}(\rho_A \log \rho_A)$, defined as the von Neumann entropy associated with the reduced density matrix $\rho_A$ of a subsystem. A close relative of the entanglement entropy is the Renyi entanglement entropy $S_n = - \frac{1}{n-1} \log \mathrm{tr}(\rho^n_A)$.

To date, the most impressive progress has been made in the study of entanglement entropy of one dimensional critical states. Here, it has been shown\cite{Holzhey,Cardy} that for systems described by conformal field theories (CFT's) the entanglement entropy of a system of length $L$ behaves as,
\beq S = \frac{c}{3} \log L/a + \gamma \label{S1d}\eeq
where $a$ is the short distance cut-off. The coefficient $c$ is a universal number known as the central charge of the CFT. The subleading constant $\gamma$ depends on the system geometry ({\it e.g.} the ratio of the subsystem size to the full system size). Although $\gamma$ is not fully universal as is clear from the cut-off dependence of (\ref{S1d}), it is universal up to an additive constant. The universal behavior of the entanglement entropy (\ref{S1d}) has proved useful for extraction of the central charge of the governing CFT in numerical density-matrix renormalization group (DMRG) studies of one-dimensional critical systems. 

Our present understanding of entanglement in dimension $d > 1$ is far less complete. However, it is generally expected that for both critical and non-critical systems the leading contribution to the entanglement entropy  scales as the area of the subsystem boundary  ${\cal A}$, $S = C {\cal A}/a^{d-1}$.\footnote{A notable exception is provided by systems with a Fermi-surface, where the area law receives a multiplicative logarithmic correction.\cite{Gioev,Wolf}} This ``area law" contribution is related to short-range entanglement in the vicinity of the boundary, and as a result, the proportionality constant $C$ is non-universal. However, for critical scale invariant systems one expects a subleading, fully universal, geometric contribution $\gamma$ to the entanglement entropy,
\beq S = C \frac{{\cal A}}{a^{d-1}} + \gamma \label{Sd}\eeq
The scaling form (\ref{Sd}) is believed to hold for scale invariant systems in $d = 2$ with arbitrary smooth subsystem boundary and in $d = 3$ with flat subsystem boundary.\cite{RyuTak1,RyuTak2,Fursaev,CasiniHuerta} Additional logarithmic contributions are expected in $d = 2$ if the boundary has sharp corners and in $d = 3$ if the boundary is curved.  
We note that all the above stated results/hypotheses on the entanglement entropy apply also to the Renyi entanglement entropy $S_n$, with the constants $C$ and $\gamma$ acquiring a dependence on $n$.

The scaling hypothesis (\ref{Sd}) relies on the following argument. The entanglement entropy, being a dimensionless quantity, can only depend on ratios of length (or energy) scales. However, in a scale invariant theory, the only two length scales are the total system size $L$ and the short distance cut-off $a$, with the corresponding energy scales $1/L^z$ and $1/a^z$, where $z$ is the dynamical critical exponent. Thus, any dependence of the entanglement entropy on the system size must come together with the dependence on the short distance cut-off. The variation $a \frac{d S}{da}$ comes from short-range entanglement in the vicinity of the boundary and is expected to take the form of some local geometric quantity integrated over the boundary area. If the boundary is straight and has no corners, there are no non-trivial local geometric quantities, and the only possible cut-off dependent contribution to the entanglement entropy is proportional to the integral of $1$ over the boundary, {\it i.e.} the boundary area, in accord with Eq.~(\ref{Sd}).  On the other hand, in two dimensions, if the boundary of the subsystem has corners, each corner can contribute a constant to $a \frac{d S}{da}$ resulting in an entanglement entropy which depends logarithmically on $a$ and hence on $L$. For a generalization of these arguments to curved boundaries, see Refs.~\onlinecite{Solodukhin,MCS,GTV}. 

In a recent breakthrough it has become possible to extract the Renyi entanglement entropy of quantum systems using Monte-Carlo simulations.\cite{Roger2010} One of the first applications of the method of Ref.~\onlinecite{Roger2010} has been to study the entanglement entropy of a spin-$1/2$ Heisenberg model on a two-dimensional square lattice.\cite{Roger2011} The Monte-Carlo simulations were performed using a torus geometry, with the subsystem being either a cylinder or a square. As expected, the leading contribution to the Renyi entanglement entropy was found to scale linearly with the system size. However, surprisingly, a subleading logarithmic correction was observed for both geometries studied. 

The ground state of the Heisenberg model on a two-dimensional square lattice is known to spontaneously break the $SU(2)$ spin-rotation symmetry to a $U(1)$ subgroup. Thus, in the infinite volume limit the ground state is infinitely degenerate and the ground state manifold is a two-dimensional sphere, whose points are labeled by the orientation of the N\'eel order parameter $\vec{n}$. The low energy excitations above a particular ground state are two linearly dispersing Goldstone bosons, known as spin-waves. The interactions between the spin-waves are irrelevant in the low-energy limit and one, thus, simply obtains a theory of two free scalar bosons. Naively, this theory is scale invariant and so according to Eq.~(\ref{Sd}), one does not expect any subleading logarithmic contributions to the entanglement entropy as long as the boundary of the subregion is smooth, {\it e.g.} for the cylindrical subregion geometry. For the square subregion geometry, one could attribute the logarithmic correction to the corners of the boundary, however, the Monte-Carlo estimate of the coefficient of the logarithm differs both in sign and by one order of magnitude compared to the previous calculation based on a free bosonic field theory.\cite{CasiniHuerta}

The caveat to the above discussion is that a system with a spontaneously broken continuous symmetry is, in fact, not scale invariant at low energy in the same way that a CFT is. The reason for this is that symmetry is always restored in a finite volume and instead of the degenerate vacuum manifold one has a unique ground state and a ``tower" of excited states carrying a definite charge under the symmetry group. For instance, in the case of the Heisenberg model with its $SU(2) \to U(1)$ symmetry breaking, the tower of states can be described by an effective Hamiltonian,
\beq H_{\mathrm{tower}} = \frac{c^2 \vec{S}^2}{2\rho_s V} \label{Htower}\eeq
where $\vec{S}$ is the total spin of the system, $c$ is the spin-wave velocity, $\rho_s$ is the spin-stiffness and $V$ is the volume of the system. Note that the spacing between the energy levels of the tower scales as $V^{-1} = L^{-d}$. For system dimension $d > 1$, where spontaneous breaking of continuous symmetry is possible, this spacing is much smaller than the spin-wave gap $c/L$.

\begin{figure}[t]
\includegraphics[width = 150mm, height = 93mm]{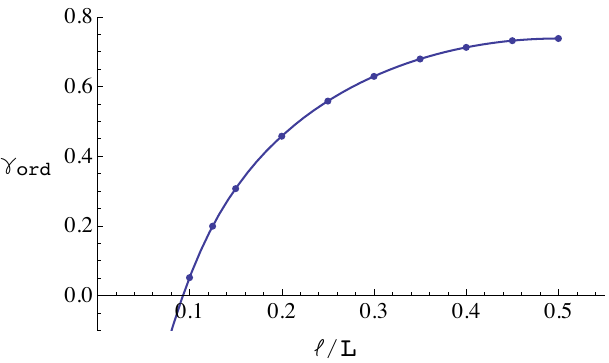}
\caption{The universal constant contribution $\gamma_{\mathrm{ord}}$, Eq.~(\ref{DeltaS}), to the Renyi entropy $S_2$ of the $O(3)$ non-linear  $\sigma$-model in dimension $d = 2$ (describing {\it e.g.} the square lattice Heisenberg model). In the geometry studied here, the total system is a torus of size $L \times L$, while the subsystem is a cylinder of size $\ell \times L$. The details of the calculation are described in section \ref{sec:eewf}. The solid line is a guide to eye. }
\label{fig:gamma}
\end{figure}

Thus, in a system with spontaneous breaking of continuous symmetry, to a given length scale $L$ there correspond two infra-red energy scales: the tower of states spacing $c^2/(\rho_s L^d)$ and the spin-wave gap $c/L$. The ratio of these energy scales $\frac{c}{\rho_s L^{d-1}}$ corresponds to the square of the magnitude of order parameter fluctuations associated with a spin-wave mode (normalized by the total size of the order parameter manifold).   In general, we expect the entanglement entropy to depend on this ratio, so by dimensional analysis
\beq S = S(L/a, \rho_s L^{d-1}/c)\eeq
Note that in the large $L$ limit, $\rho_s L^{d-1}/c$ is much greater than one.\footnote{Clearly, for $L \to \infty$, the quantity $\rho_s L^{d-1}/c \gg 1$. One can, however, ask if this quantity can become of $O(1)$ or smaller in a system close to a continuous phase transition into an antiferromagnetically disordered phase where $\rho_s \to 0$. For the effective low-energy description of the antiferromagnet to apply, we need the system size $L$ to be much larger than the correlation length of the system $\xi$, thus, $\rho_s L^{d-1}/c \gg \rho_s \xi^{d-1}/c$. A priori, there need not be a relation between $\rho_s$ and $\xi$. However, for all quantum phase transitions known to the authors, $\rho_s \xi^{d-1}/c$ either goes to a constant at the transition or diverges.} In this paper, we demonstrate that for a system with $O(N) \to O(N-1)$ symmetry breaking, for a smooth subsystem boundary in $d = 2$ (straight boundary in $d = 3$),
\beq S = C \frac{\cal A}{a^{d-1}}  + b \log (\rho_s L^{d-1}/c) + \gamma_{\mathrm{ord}}, \quad b = \frac{N-1}{2}\label{DeltaS}\eeq
Thus, the entanglement entropy contains an extra subleading term which diverges logarithmically with the system size.  The coefficient of this term $b$ is directly expressed in terms of the number of Goldstone modes $N-1$. In particular, for the case of the Heisenberg antiferromagnet $N = 3$, $b = 1$. We note that in addition to the logarithmic correction in Eq.~(\ref{DeltaS}) there also appears a finite constant $\gamma_{\mathrm{ord}}$ that depends on the system geometry. Unlike in the case of entanglement entropy in one dimension, Eq.~(\ref{S1d}), where the presence of a logarithm rendered $\gamma$ universal only up to an additive contribution, here $\gamma_{\mathrm{ord}}$ is fully universal, as all the short-distance physics is absorbed into the order-parameter stiffness $\rho_s$ and the Goldstone velocity $c$. Given a subsystem geometry, our calculation method allows us to determine  $\gamma_{\mathrm{ord}}$ numerically; in Fig.~\ref{fig:gamma} we present the result for the geometry studied in the Monte-Carlo simulations of Ref.~\onlinecite{Roger2011, Rogertoappear}.

We also extend the result (\ref{DeltaS}) to the case when $d = 2$ and the boundary has corners, or $d  =3$ and the boundary is curved. Here we obtain,
\beq S = C \frac{\cal A}{a^{d-1}}  + b' \log L/a +  b \log (\rho_s L^{d-1}/c) + \gamma_{\mathrm{ord}}, \quad b = \frac{N-1}{2}\label{DeltaScorn}\eeq  
The coefficient $b'$ can be computed in a free theory of Goldstone fields. For instance, for the $d = 2$ case with corners,
\beq b' = (N-1) \sum_i b_{\mathrm{corn}}(\varphi_i) \label{eq:bpcorn}\eeq
The sum in Eq.~(\ref{eq:bpcorn}) is over the corners of the boundary; $b_{\mathrm{corn}}(\varphi_i)$ denotes the corner log coefficient in a free scalar bosonic theory. This coefficient depends on the corner angle $\varphi_i$ and has been computed in Ref.~\onlinecite{CasiniHuerta}. 


In addition to the entanglement entropy, we compute the Renyi entropies $S_n$, which are also found to satisfy the scaling forms in Eqs.~(\ref{DeltaS}), (\ref{DeltaScorn}). The coefficient of the area law $C$  now depends on the replica index $n$, so does the finite size constant $\gamma_{\mathrm{ord}}$ and the corner/curvature coefficient $b'$. However, the coefficient of the logarithmic correction $b$ is found to be independent of the replica index.

We also study the full low energy spectrum of the entanglement Hamiltonian $H_E$ defined in terms of the reduced density matrix of the subsystem $\rho_A$ as $\rho_A = \exp(-H_E)$. We find,
\beq H_E = \frac{\vec{S}^2_A}{2 I} + \sum_\epsilon \epsilon\, a^{\dagger}_\epsilon a_{\epsilon} \label{eq:spectrumintro}\eeq
Here $\vec{S}_A$ denotes the $O(N)$ spin of the subsystem and $a_\epsilon$, $a^{\dagger}_\epsilon$ are bosonic annihilation/creation operators. The first term in Eq.~(\ref{eq:spectrumintro}) has the same form as the tower of states sector (\ref{Htower}) of the physical Hamiltonian, while the second term is harmonic. The coefficient $I$ to logarithmic accuracy is given by
\beq I = \frac{\rho_s}{2 \pi} {\cal A} \log(L/a)  \label{eq:Iintro}\eeq
The gaps $\epsilon$ in the harmonic sector are found to scale as $(\log L/a)^{-1}$. Thus, in $d > 1$ where spontaneous breaking of continuous symmetry is possible, the entanglement gap in the tower of states sector $I^{-1} \sim (L^{d-1} \log L/a)^{-1}$ is parametrically smaller than the gap in the harmonic sector. 

 We note that recent DMRG studies of the entanglement spectrum in the superfluid phase of the 2d bosonic Hubbard model are in reasonable agreement with our analytical result (\ref{eq:spectrumintro}).\cite{EntSpectHubbard} In particular, Ref.~\onlinecite{EntSpectHubbard}  has observed a tower of states structure of the low-lying entanglement spectrum with the entanglement gap scaling inversely with the length of the subsystem boundary, as predicted in Eq.~(\ref{eq:Iintro}). Note that in a superfluid, the system displays spontaneous breaking of $U(1) \equiv O(2)$ symmetry, so the role of $\vec{S}_A$ is played by the subsystem particle number $N_A$ (more precisely, its deviation from the ground state expectation value $\delta N_A = N_A - \langle N_A \rangle$). The higher lying states in the entanglement spectrum were found in Ref.~\onlinecite{EntSpectHubbard} to exhibit a much weaker system size dependence, consistent with our result $\epsilon \sim (\log L/a)^{-1}$.

For $O(2)$ symmetry, the result (\ref{eq:spectrumintro}) also applies to 1d superfluids/XY magnets, which are described by Luttinger liquid theory. For this 1d case, the tower of states gap and the harmonic gap in the entanglement spectrum scale in the same way as $(\log L/a)^{-1}$. This is not surprising:  Luttinger liquid theory is a CFT, and a powerful result on 1d CFTs states that the entanglement spectrum of a segment of length $\ell$ embedded in an infinite line is identical to the physical spectrum of the system on an open segment of length $\ell_{strip} = \frac{1}{\pi} \log \ell/a$.\cite{Holzhey} Thus, for $d = 1$, the first term in Eq.~(\ref{eq:spectrumintro}) is the tower of states part of the physical spectrum on an open segment and the second term is the spin-wave part of the physical spectrum on an open segment. We stress that such direct full correspondence of physical spectrum and entanglement spectrum is exact only in $d = 1$. We also note that recent DMRG simulations of Ref.~\onlinecite{EntSpectCFT} have provided a numerical confirmation of this correspondence for a number of systems described by 1d CFTs, including the superfluid phase of the 1d Bose-Hubbard model and the Luttinger-liquid phase of the XXZ chain.


The results of the present paper are based on the analysis of the $O(N)$ non-linear $\sigma$-model, which provides the full low-energy description of the system. We find the ground state wave-function for the non-linear $\sigma$-model in a finite volume, compute the reduced density matrix $\rho_A$ and obtain its spectrum, Eq.~(\ref{eq:spectrumintro}). From this spectrum, we then compute the entanglement entropy. This procedure is very similar to the correlation matrix technique used to find the entanglement spectrum and entanglement entropy in a free bosonic field theory,\cite{Peschel} except we pay special attention to the compact nature of the order parameter manifold. We also separately confirm the result on entanglement entropy (\ref{DeltaS}) by performing replica method calculations on the non-linear $\sigma$-model in the path-integral formulation (see Appendix). We stress that all our results are exact in the large system limit.

Curiously, to get an intuition about our exact results it is sufficient to consider a simple quantum mechanical model of two coupled quantum $O(N)$ rotors $\vec{n}_A$ and $\vec{n}_B$, representing the average order parameter in subsystem $A$ and its complement $B$,
\beq H = \frac{c^2 \vec{L}^2_A}{2 \rho_s V_A} + \frac{c^2 \vec{L}^2_B}{2 \rho_s V_B}  - J \vec{n}_A \cdot \vec{n}_B \label{Hrotor}\eeq
Here $\vec{L}_{A,B}$ is the angular momentum of each rotor. $V_A$, $V_B$ and $V$ denote the volumes of each subsystem and of the total system respectively. We choose the coupling $J \sim \rho_s L^{d-2}$ appropriately to reflect the finite order-parameter stiffness of the system. One finds that the entanglement Hamiltonian corresponding to the ground state wave-function of this model, indeed, has a tower of states form and reproduces the logarithmic correction to the entanglement entropy in Eq.~(\ref{DeltaS}).

We would like to note that the presence of logarithmic corrections to the entanglement entropy of a Heisenberg antiferromagnet has been theoretically pointed out previously in Ref.~\onlinecite{Karyn}. The authors of Ref.~\onlinecite{Karyn} have used the spin-wave (large $S$) expansion of the Heisenberg model. The effect of symmetry restoration in a finite volume has been mimicked by an application of a staggered magnetic field $h \sim 1/V^2$, suitably chosen to give a zero net staggered moment. The final step of the calculation of the entanglement entropy has been performed numerically and finite size scaling was used to extract the coefficient of the logarithmic divergence $b \approx 0.93$. This value is quite close to our exact result $b = 1$. Furthermore, we have repeated the calculation in Ref.~\onlinecite{Karyn} for system sizes up to $100 \times 200$ sites and found that $b$ approaches unity (within a percent) as the system size is increased.\footnote{We also
note that the calculation in Ref.~\onlinecite{Karyn} can be reformulated using the Schwinger boson representation of spin operators.  This yields an $SU(2)$ spin-symmetric mean-field ansatz wavefunction for the antiferromagnetic state that breaks the $SU(2)$ symmetry spontaneously in the thermodynamic limit.\cite{auerbach}} While our calculation of the coefficient $b$ agrees with the semi-numerical method of Ref.~\onlinecite{Karyn}, our approach has the advantage of being exact: in particular, it correctly treats the compactness of the order parameter and does not rely on the $1/S$ expansion.    

An attempt to make a connection between the restoration of symmetry in a finite volume and the appearance of subleading logarithmic terms in the entanglement entropy has also been made in Ref.~\onlinecite{Roger2011}. Here the authors started with a mean-field N\'eel state of a Heisenberg antiferromagnet and averaged it over all orientations of the order parameter. The entanglement entropy of the resulting spin-singlet state was found to be $S = \log {\cal N} = d \log(L/a)$, where ${\cal N}$ is the total number of lattice sites (here and below we drop the constant piece in $S$). However, this approximation is too crude: it gives the coefficient of the logarithmic divergence to be $b = d/(d-1)$, instead of our exact result $b = 1$. The physical reason why the approach of Ref.~\onlinecite{Roger2011} fails is that it does not take into account the existence of spin-waves, while as we have argued above, the presence of both the tower of states and spin-waves is crucial for the logarithmic correction to entanglement entropy in Eq.~(\ref{DeltaS}). The estimate of Ref.~\onlinecite{Roger2011} for an antiferromagnet is physically similar to the exact result one obtains in the case of a free Bose gas where $S = \frac{1}{2} \log {\cal N} = \frac{d}{2} \log{L/a}$, with ${\cal N}$ - the total number of particles and $a$ - the average interparticle distance. On the other hand,  for an interacting Bose gas, {\it i.e.} a superfluid, our result, Eq.~(\ref{DeltaS}), with $N= 2$ gives a logarithmic correction $\Delta S = \frac{d-1}{2} \log \left((\rho_s/c)^{1/(d-1)} L\right)$. Thus, based on entanglement entropy we can distinguish a free Bose gas, where no Goldstone boson is present, from an interacting superfluid, which has a Goldstone mode.

This paper is organized as follows. As a warm-up, we begin by calculating the entanglement properties of the toy rotor model (\ref{Hrotor}) in section \ref{sec:rotor}. In section \ref{sec:wf}, we determine the ground state wave-function of the $O(N)$ non-linear $\sigma$-model and use it to compute the entaglement spectrum and the entanglement entropy. Some concluding remarks are presented in section \ref{sec:concl}. In the appendix, we give an alternative calculation of the entanglement entropy in the $O(N)$ non-linear $\sigma$-model using the replica method and show that the result is in complete agreement with of our wave-function calculation in section \ref{sec:wf}.  

\section{Warm up: rotor model.}
\label{sec:rotor}
In this section, we calculate the Renyi entropy in the model (\ref{Hrotor}), describing two quantum $O(N)$ rotors $\vec{n}_A$ and $\vec{n}_B$ of unit length. These rotors are taken to represent the average orientation of the order parameter in subsystems $A$ and $B$. We choose the coupling $J$ between the rotors in the following way. Recall that when the system is placed in a box of size $L$, the energy cost to twist the order parameter by an angle $\theta$ between the two sides of the box is related to the order-parameter stiffness $\rho_s$ via,
\beq \rho_s = \frac{1}{L^{d-2}} \frac{\d^2 E}{\d \theta^2}\bigg|_{\theta = 0} \eeq
Thus, we take $J \sim \rho_s L^{d-2}$. Note that this value of $J$ is approximate and should be understood in a scaling sense only.  

To find the spectrum of the model (\ref{Hrotor}) it is convenient to first work with the Lagrangian formulation,
\beq L = \frac{\rho_s V_A (\d_{\tau} \vec{n}_A)^2} {2 c^2} + \frac{\rho_s V_B (\d_{\tau} \vec{n}_B)^2}{2 c^2}  - J \vec{n}_A \cdot \vec{n}_B \label{Lrotor}\eeq
Let us introduce the average and relative coordinates $\vec{n}$ and $\delta n_\alpha$ via,
\bea \vec{n}_A &=& \vec{n} \sqrt{1 - a^2 (\delta n_\alpha)^2} + a \vec{e}_\alpha \delta n_\alpha , \quad a = V_B/(V_A + V_B)\nn\\
\vec{n}_B &=& \vec{n} \sqrt{1 - b^2 (\delta n_\alpha)^2} + b \vec{e}_\alpha \delta n_\alpha , \quad b = -V_A/(V_A + V_B) \label{coordchange}\eea
Here, $\vec{n}^2 = 1$, and $\vec{e}_\alpha$ are $(N-1)$ unit vectors forming an orthonormal set together with $\vec{n}$. We expect the fluctuations of the average coordinate $\vec{n}$ to be parametrically slower than those of the relative coordinate $\delta n$. Moreover, we expect the fluctuations of $\delta n$ to be small. Therefore, expanding the Lagrangian (\ref{Lrotor}) to leading order in $\delta n$ and in derivatives of $\vec{n}$ we obtain,
\bea L &\approx& L_{n} + L_{\delta n} \label{L}\\
L_{n} &=& \frac{\rho_s V}{2 c^2} (\d_\tau \vec{n})^2\label{Lavg}\\
L_{\delta n} &=& \frac{\rho_s V_r}{2 c^2} (\d_\tau \delta n_\alpha)^2 + \frac{J}{2} (\delta n_\alpha)^2 \label{Lrel}\eea
Here $V = V_A + V_B$ is the total volume and $V_r = V_A V_B/(V_A + V_B)$ is the reduced volume. We see that the Lagrangians of the average and relative coordinates decouple. The dynamics of the average coordinate $\vec{n}$ are governed by the quantum rotor Lagrangian (\ref{Lavg}); the associated Hamiltonian
\beq H_{n} = \frac{c^2 \vec{L}^2_{\mathrm{total}}}{2 \rho_s V}, \quad \vec{L}_{\mathrm{total}} = \vec{L}_A + \vec{L}_B, \label{Havg}\eeq
is, indeed, the appropriate ``tower of states" Hamiltonian describing the lowest energy excitations of a system with spontaneous symmetry breaking. 
Here $L_{a b} = -i \left( n^a  \frac{\d}{\d n^b} - n^b  \frac{\d}{\d n^a}\right)$, $a, b = 1\ldots N$, are angular momentum operators, i.e. generators of $O(N)$ rotations of $\vec{n}$, and hence of $\vec{n}_A, \vec{n}_B$. They can, thus, be identified with the total $O(N)$ angular momentum (spin) of the system. We use the short-hand $\vec{L}^2_{\mathrm{total}} = \sum_{a < b} L^2_{ab}$, which is also equal to the Laplacian, $-\nabla^2$, on the $\vec{n}$ sphere. 

The dynamics of the relative coordinate $\delta n$ are governed by the Lagrangian (\ref{Lrel}) describing an $N-1$ dimensional harmonic oscillator. The frequency of this harmonic oscillator is given by
\beq \omega = \sqrt{\frac{c^2 J}{\rho_s V_r}} \sim \frac{c}{L} \label{omegarel}\eeq
In the true physical system, the spectrum of the ``relative motion" involves $N-1$ Goldstone modes with a dispersion $\omega = c |\vec{k}|$, where the momentum $\vec{k}$ is quantized in a finite geometry. Our rotor model (\ref{Hrotor}) replaces this multi-mode spectrum with a single $N-1$ dimensional oscillator whose energy (\ref{omegarel}) is of order of the finite-size gap of the Goldstone modes. It turns out that such a replacement is sufficient for capturing the logarithmic contribution to the entanglement entropy in Eq.~(\ref{DeltaS}). 

The ground-state wave-function corresponding to the Lagrangian (\ref{L}) is a product of a $\vec{L}_{\mathrm{total}} = 0$ wave-function of the rotor Hamiltonian (\ref{Havg}) and the ground state wave-function of the harmonic oscillator (\ref{Lrel}),
\bea \psi(\vec{n}, \delta n) &=& \frac{1}{\sqrt{|S^{N-1}|}}\frac{1}{(\pi \xi^2)^{(N-1)/4}} \exp\left(-(\delta n_\alpha)^2/(2 \xi^2)\right) \nn\\ &\approx& \frac{1}{\sqrt{|S^{N-1}|}}\frac{1}{(\pi \xi^2)^{(N-1)/4}} \exp\left(-(\vec{n}_A - \vec{n}_B)^2/(2 \xi^2)\right) \label{psindn}\eea
Here, $|S^{N-1}| = 2 \pi^{N/2}/\Gamma(N/2)$ is the volume of a unit sphere $S^{N-1}$ and  
\beq \xi = \left(\frac{c^2}{\rho_s V_r J}\right)^{1/4} \sim \left(\frac{c}{\rho_s L^{d-1}}\right)^{1/2} \ll 1 \label{eq:xi}\eeq
Note that the condition $\xi \ll 1$ guarantees that the amplitude of the relative fluctuations is, indeed, small.

We proceed to compute the reduced density matrix $\rho_A$ from the wave-function (\ref{psindn}),
\beq \rho_A(\vec{n}_A, \vec{n}'_A) = \int d \vec{n}_B\, \psi(\vec{n}_A, \vec{n}_B)\psi^*(\vec{n}'_A, \vec{n}_B) \label{rhoA1}\eeq
The contributions to the integral over $\vec{n}_B$ above come from $|\vec{n}_A - \vec{n}_B|, |\vec{n'}_A - \vec{n}_B| \sim \xi \ll 1$. Therefore, $\rho_A(\vec{n}_A, \vec{n}'_A)$ is non-negligible only for $|\vec{n}_A - \vec{n}'_A| \ll 1$. We may change variables in Eq.~(\ref{rhoA1}) to $\delta \vec{n} = \vec{n}_A - \vec{n}_B$ and, to leading order, take both $\delta \vec{n}$ and $\vec{n}_A - \vec{n}'_A$ to lie in the tangent plane of $\vec{n}_A$. The integral over $\delta \vec{n}$  then becomes Gaussian and gives,
\beq \rho_A(\vec{n}_A, \vec{n}'_A) = \frac{1}{|S^{N-1}|} \exp\left(-(\vec{n}_A - \vec{n}'_A)/(4 \xi^2)\right)\eeq
Since $\vec{L}^2 = - \nabla^2$ is just the Laplacian on the sphere, we may use the heat kernel expansion
\beq \langle \vec{n}| e^{-s \vec{L}^2}| \vec{n}'\rangle \to \frac{1}{(4 \pi s)^{(N-1)/2}} \exp(-(\vec{n} - \vec{n}')^2/(4s)), \quad s \to 0 \label{HeatKer}\eeq
Thus, as $\xi \ll 1$, we may write,
\beq \rho_A \approx \frac{(4 \pi \xi^2)^{(N-1)/2}}{|S^{N-1}|} e^{- \xi^2 \vec{L}^2_A} \label{rhoL2}\eeq
Defining the entanglement Hamiltonian $H_E$ as $\rho_A = \exp(-H_E)$, we obtain,
\beq H_E = \xi^2 \vec{L}^2_A + const \label{HErot}\eeq
Hence, the entanglement Hamiltonian has the same form as the physical tower of states Hamiltonian (\ref{Havg}). The entanglement gap in Eq.~(\ref{HErot}) is found to scale as $\xi^{2} \sim \frac{c}{\rho_s L^{d-1}}$. In the next section we will perform an analysis of the full non-linear $\sigma$-model, instead of the toy rotor model studied here, revealing that the lowest branch of the exact entanglement Hamiltonian, indeed,  has a tower of states form $H_E = \frac{\vec{L}^2_A}{2 I}$, with $I$ given by Eq.~(\ref{eq:Iintro}). Thus, up to a logarithmic factor, the scaling of the entanglement gap in Eq.~(\ref{HErot}) agrees with the exact result (\ref{eq:Iintro}). Note that $\xi$ in Eq.~(\ref{HErot}) is determined via Eq.~(\ref{eq:xi}) by the coupling $J$ of the rotor model that we simply postulated to scale as $J \sim 
\rho_s L^{d-1}$. By choosing $J$ appropriately, we can force the tower of states spectrum (\ref{HErot}) to match the exact result (\ref{eq:Iintro}).

We can now compute the Renyi entropy of the rotor model. From Eq.~(\ref{rhoL2})
\beq \mathrm{tr}(\rho^n_A) \approx \frac{(4 \pi \xi^2)^{n(N-1)/2}}{|S^{N-1}|^n} \mathrm{tr}(e^{- n \xi^2 L^2_A}) \approx \frac{ (4 \pi \xi^2)^{(n-1)(N-1)/2}}{|S^{N-1}|^{n-1} n^{(N-1)/2}}\eeq
where we've used Eq.~(\ref{HeatKer}) in the last step. Therefore, 
\bea S_n &=& -\frac{1}{n-1} \log \mathrm{tr}(\rho^n_A) = \frac{N-1}{2}\left(\log \frac{1}{4 \pi \xi^2} + \frac{\log n}{n-1} \right) + \log |S^{N-1}| \label{RenyiOrdconst}\\ &\sim& \frac{N-1}{2} \log \frac{\rho_s L^{d-1}}{c}\label{RenyiOrd}\eea
Note that if we wish to interpret the result (\ref{RenyiOrd}) in terms of the actual physical system rather than the toy rotor model, we must remember that Eq.~(\ref{RenyiOrd}) only captures the contribution of the lowest branch of the entanglement spectrum given by Eq.~(\ref{HErot}). There will also be a contribution from the higher lying part of the entanglement spectrum. In the next section, we will determine this higher lying part, and show that it contributes a standard area law term to the Renyi entropy, so we identify the Renyi entropy of the rotor model, Eq.~(\ref{RenyiOrd}), with the logarithmic correction in Eq.~(\ref{DeltaS}). In addition, since the result (\ref{RenyiOrdconst}) depends on $\xi$ ($J$), and hence on the details of the spectrum of ``relative fluctuations" (Goldstone modes), it is only logarithmically accurate. In particular, it does not capture the universal geometric constant $\gamma_{\mathrm{ord}}$ of Eq.~(\ref{DeltaS}).\footnote{If we choose $J$ to reproduce the exact gap in the tower of states sector of the entanglement spectrum (\ref{eq:Iintro}), strictly speaking, Eq.~(\ref{RenyiOrd}) will also contain a $\log \log L/a$ term. It turns out that this term combines with the contribution from the higher lying modes in the entanglement spectrum to give an overall area-law contribution.} However, our full calculation in the next section will allow us to determine $\gamma_{\mathrm{ord}}$, as well.

\section{Wave-function method.}
\label{sec:wf}
In this section, we calculate the entanglement entropy in the $O(N)$ non-linear $\sigma$-model using the wave-function method. The action of the theory is given by,
\beq S = \frac{\rho_s}{2} \int d^d x  d\tau \left( \frac{1}{c^2} (\d_\tau \vec{n})^2 + (\nabla \vec{n})^2\right) \label{eq:Nlsmwf}\eeq
with $n^a$, $a = 1 \ldots N$,  an $N$-dimensional unit vector. We will set the spin-wave velocity $c = 1$ below and restore it in the final results. For simplicity, we consider the theory on a spatial torus, although a system with open boundaries can be treated with minimal modifications.

Following the standard treatment, we write
\beq \vec{n}( \vec{x}, \tau) = \vec{n}_0(\tau)  \sqrt{1 - \frac{\tilde{\pi}_\alpha(\vec{x}, \tau) \tilde{\pi}_\alpha(\vec{x}, \tau)}{\rho_s}} + \frac{\vec{e}_\alpha(\tau) {\tilde{\pi}_\alpha(\vec{x}, \tau)}}{\sqrt{\rho_s}} \label{eq:ntildepi} \eeq
Here, $\alpha = 1\ldots N-1$. $\vec{n}_0$ and $\vec{e}_\alpha$ are unit vectors that together form an orthonormal basis: $\vec{n}_0 \cdot \vec{e}_\alpha = 0$, $\vec{e}_\alpha \cdot \vec{e}_\beta = \delta_{\alpha \beta}$. The fields $\tilde{\pi}_\alpha$ are constrained to satisfy,
\beq \int d^d x \,\tilde{\pi}_\alpha(\vec{x}, \tau) = 0 \label{eq:piconst}\eeq
The vector $\vec{n}_0(\tau)$ describes the (slow) fluctuations of the overall direction of the order parameter in the system, while $\tilde{\pi}_\alpha(\vec{x}, \tau)$ describe the spin-wave fluctuations. Expanding the action (\ref{eq:Nlsmwf}) to leading order in $\tilde{\pi}_\alpha$, we obtain
\beq S = \frac{\rho_s V}{2} \int d\tau (\d_\tau \vec{n}_0)^2 + \frac{1}{2} \int d^d x d \tau \left( (\d_{\tau} \tilde{\pi}_{\alpha})^2+ (\nabla\tilde{\pi}_{\alpha})^2\right) \label{eq:Spi} \eeq 
where $V$ is the volume of the system.  We remind the reader that higher order terms in the expansion in $\tilde{\pi}_\alpha$ are irrelevant in the RG sense for $d > 1$. Thus, to leading order the motion of $\vec{n}_0$ and $\tilde{\pi}_\alpha$ decouples. The Hamiltonian corresponding to the action (\ref{eq:Spi}) is given by,
\beq H = H_{\mathrm{tower}} + H_{\mathrm{sw}} \label{eq:H}\eeq
where $H_{\mathrm{tower}}$ desribes the $\vec{n}_0$ sector and $H_{\mathrm{sw}}$ - the $\tilde{\pi}_{\alpha}$ sector. The action for $\vec{n}_0$ is that of a particle on a unit sphere, so
\beq H_{\mathrm{tower}} = \frac{\vec{L}^2}{2 \rho_s V} \label{eq:Htowern0}\eeq
Here $L_{a b} = -i \left( n^a_0  \frac{\d}{\d n^b_0} - n^b_0  \frac{\d}{\d n^a_0}\right)$ are angular momentum operators, i.e. generators of $O(N)$ rotations of $\vec{n}_0$, and hence of $\vec{n}$. They can, thus, be identified with the total $O(N)$ angular momentum (spin) of the system. We use the short-hand $\vec{L}^2 = \sum_{a < b} L^2_{ab}$, which is also equal to the Laplacian, $-\nabla^2$, on the $\vec{n}_0$ sphere. 

The action for $\tilde{\pi}_{\alpha}$ is quadratic, so the spin-wave Hamiltonian takes the form
\beq H_{\mathrm{sw}} = \sum_{\vec{k}\neq0 } |\vec{k}| a^{\dagger}_{\vec{k} \alpha} a_{\vec{k} \alpha} \label{eq:Hsw}\eeq
where 
\beq \tilde{\pi}_{\alpha}(\vec{x}) = \frac{1}{V^{1/2}} \sum_{\vec{k} \neq 0} \frac{1}{(2 |\vec{k}|)^{1/2}} \left(a_{\vec{k},\alpha} e^{i \vec{k} \cdot \vec{x}} + a^{\dagger}_{\vec{k},\alpha} e^{-i \vec{k} \cdot \vec{x}} \right) \label{eq:pisw}\eeq
and $\vec{k}$ are the momenta allowed by the periodic boundary conditions on the torus. Note that the $|\vec{k}| = 0$ mode is missing from the sums (\ref{eq:Hsw}), (\ref{eq:pisw}) due to the constraint (\ref{eq:piconst}). For future reference, we introduce the static propagator,
\beq \langle \tilde{\pi}_{\alpha}(\vec{x}) \tilde{\pi}_{\beta}(\vec{y}) \rangle = \delta_{\alpha \beta} \tilde{D}(\vec{x}, \vec{y}) \label{eq:pipi}\eeq
with 
\beq \tilde{D}(\vec{x}, \vec{y}) = \frac{1}{V} \sum_{\vec{k} \neq 0} \frac{1}{2 |\vec{k}|} e^{i \vec{k} \cdot (\vec{x} - \vec{y})}  \label{eq:Dtilde}\eeq

Let us now write the ground-state wave-function(al) of the system, $\psi[\vec{n}(\vec{x})]$. It is given by a product of ground state wave-functions in the $\vec{n}_0$ and $\tilde{\pi}_\alpha$ sectors. The ground state in the $\vec{n}_0$ sector carries zero angular momentum and the corresponding wave-function $\psi(\vec{n}_0) $ is a constant. The ground state wave-function in the $\tilde{\pi}_\alpha$ sector is a Gaussian; its form can be easily deduced by requiring that the wave-function reproduce the correlator (\ref{eq:pipi}). We, thus, obtain
\beq \psi[\tilde{\pi}_\alpha] \propto \exp\left(-\frac{1}{4} \int d^d x d^d y \, \tilde{\pi}_\alpha(\vec{x}) \tilde{Q}(\vec{x}, \vec{y}) \tilde{\pi}_\alpha(\vec{y})\right) \label{eq:psipi}\eeq
where
\beq \int d^d z \, \tilde{Q}(\vec{x}, \vec{z}) \tilde{D}(\vec{z}, \vec{y}) = \delta^d(\vec{x} - \vec{y}) - \frac{1}{V} \label{eq:QD} \eeq
i.e.
\beq \tilde{Q}(\vec{x}, \vec{y}) = \frac{1}{V} \sum_{\vec{k} \neq 0} {2 |\vec{k}|} e^{i \vec{k} \cdot (\vec{x} - \vec{y})} = -4 \,\nabla^2 \tilde{D}(\vec{x}, \vec{y}) \label{eq:Qexpr}\eeq  
In Eq.~(\ref{eq:psipi}) and below, we ignore the normalization of the wave-function as it will not be essential for our purposes. Thus, the overall ground state wave-function of the system is,
\bea \psi[\vec{n}] &=& \psi(\vec{n}_0) \psi[\tilde{\pi}_\alpha] \propto \exp\left(-\frac{1}{4} \int d^d x d^d y \, \tilde{\pi}_\alpha(\vec{x}) \tilde{Q}(\vec{x}, \vec{y}) \tilde{\pi}_\alpha(\vec{y})\right) \nn\\
&\approx&  \exp\left(-\frac{\rho_s}{4} \int d^d x d^d y \, n^a(\vec{x}) \tilde{Q}(\vec{x}, \vec{y}) n^a(\vec{y})\right) \label{eq:psipi2}\eea
where the last equality holds to leading order in $\tilde{\pi}_\alpha$. 

Next, we divide our system into region $A$ and its complement $B$ and compute the reduced density matrix $\rho_A$ associated with region $A$ (at this point we keep the shape of regions $A$ and $B$ arbitrary). 
\beq \rho_A(\vec{n}_A, \vec{n}'_A) = \int D \vec{n}_B(\vec{x}) \, \psi(\vec{n}_A, \vec{n}_B) \psi^*(\vec{n}'_A, \vec{n}_B) \label{eq:rhoA}\eeq  
Here, $\vec{n}_{A}(\vec{x})$ and $\vec{n}_B(\vec{x})$ denote the values of $\vec{n}(\vec{x})$ restricted to regions $A$ and $B$, respectively. Recall that we are considering configurations of $\vec{n}(\vec{x})$ with small, smooth flucutations $\tilde{\pi}_\alpha(\vec{x})$ about a global direction $\vec{n}_0$, i.e. each $\vec{n}(\vec{x})$ maps the  space into some small patch of the sphere. Since the values of $\vec{n}_B$ in the two wave-functions in the integrand of Eq.~(\ref{eq:rhoA}) are identified, we can also take $\vec{n}_A(\vec{x})$, $\vec{n}'_A(\vec{x})$, as well as $\vec{n}_B(\vec{x})$ to lie in the same small patch of the sphere. We then compute $\rho_A(\vec{n}_A, \vec{n}'_A)$ patch by patch. By rotational symmetry, it is sufficient to compute $\rho_A(\vec{n}_A, \vec{n}'_A)$ in the patch centered at the North pole. Then, write
\beq \vec{n}(\vec{x}) = (\pi_a(\vec{x})/\sqrt{\rho_s}, \sqrt{1 - \pi_{a}(\vec{x}) \pi_{a}(\vec{x})/\rho_s}) \label{eq:NorthPole}\eeq 
Here, the index $a$ on $\pi_a(\vec{x})$ runs over $1\ldots N-1$, and unlike $\tilde{\pi}_\alpha$, $\pi_a(\vec{x})$ is {\it unconstrained}. Expanding $\rho_A$ to leading order in $\pi$,
\beq \rho_A(\vec{\pi}_A, \vec{\pi}'_A) \propto \int D \pi_B(\vec{x}) \exp\left[-\frac{1}{4} \left(\vec{\pi}^T_A \tilde{Q}_{AA} \vec{\pi}_A  + {\vec{\pi}'^T_A} \tilde{Q}_{AA} \vec{\pi}'_A + 2 \vec{\pi}^{T}_B \tilde{Q}_{B B} \vec{\pi}_B + 
2 (\vec{\pi}_A + \vec{\pi}'_A)^T \tilde{Q}_{AB} \vec{\pi}_B \right)\right] \label{eq:rhoApi}\eeq
Here and below we use the short-hand notation $v^T O w = \int d^d x  d^d y\,  v(\vec{x}) O(\vec{x}, \vec{y}) w(\vec{y})$. $\vec{\pi}_A$ and $\vec{\pi}_B$ denote $\vec{\pi}(\vec{x})$ restricted to regions $A$ and $B$, respectively. Similarly, $\tilde{Q}_{AA}$ denotes $\tilde{Q}(\vec{x}, \vec{y})$ with both arguments restricted to region $A$, $\tilde{Q}_{BB}$ denotes $\tilde{Q}(\vec{x}, \vec{y})$ with both arguments restricted to region $B$, and $\tilde{Q}_{AB}$ denotes $\tilde{Q}(\vec{x}, \vec{y})$ with $\vec{x}$ restricted to region $A$ and $\vec{y}$ restricted to region $B$. Performing the integral over $\pi_B$ in Eq.~(\ref{eq:rhoApi}),
\beq \rho_A(\vec{\pi}_A, \vec{\pi}'_A) \propto  \exp\left[-\frac{1}{8} \left(2 \vec{\pi}^T_A \tilde{Q}_{AA} \vec{\pi}_A  + 2 {\vec{\pi}'^T_A} \tilde{Q}_{AA} \vec{\pi}'_A -  
 (\vec{\pi}_A + \vec{\pi}'_A)^T \tilde{Q}_{AB} \tilde{Q}^{-1}_{BB} \tilde{Q}_{BA}  (\vec{\pi}_A + \vec{\pi}'_A) \right)\right]\eeq
where $\tilde{Q}^{-1}_{BB}$ is defined with its arguments over region $B$ and satisfies,
\beq \int_B d^d z \, \tilde{Q}^{-1}_{BB}(\vec{x}, \vec{z}) \tilde{Q}_{BB} (\vec{z}, \vec{y}) = \delta^d(\vec{x}-\vec{y}) \eeq
Note that although $\tilde{Q}$ defined over the entire system does not have an inverse due to the presence of the $\vec{k} =0$ zero mode, $\tilde{Q}_{BB}$ - the restriction of $\tilde{Q}$ to region $B$ generally does possess an inverse. Now from Eq.~(\ref{eq:QD}) we deduce,
\beq \tilde{Q}_{AB} \tilde{Q}^{-1}_{BB} \tilde{Q}_{BA} = \tilde{Q}_{AA} - \tilde{D}^{-1}_{AA} + \frac{\tilde{D}^{-1}_{AA} P_A \tilde{D}^{-1}_{AA}}{\mathrm{tr}(P_A \tilde{ D}^{-1}_{AA})} \label{eq:BAB}\eeq 
where $\tilde{D}^{-1}_{AA}$ is defined with its arguments over region $A$ and satisfies,
\beq \int_A d^d z \tilde{D}^{-1}_{AA}(\vec{x}, \vec{z}) \tilde{D}_{AA}(\vec{z},\vec{y}) = \delta^d(\vec{x}-\vec{y})\eeq
Again, even though $\tilde{D}$ defined over the entire system does not possess an inverse, $\tilde{D}_{AA}$ generally does. $P_A$ is also defined with its arguments over region $A$ and is given by, $P_A(\vec{x}, \vec{y}) = \frac{1}{V_A}$, with $V_A$ - the volume of region $A$. In other words, $P_A$ is the projector onto the constant mode $v_0(\vec{x}) = \frac{1}{\sqrt{V_A}}$ over region $A$. Now using Eq.~(\ref{eq:BAB}), $\rho_A$ takes the form,
 \bea \rho_A(\vec{\pi}_A, \vec{\pi}'_A) &\propto&  \exp\left(-\frac{1}{8} (\vec{\pi}_A - \vec{\pi}'_A)^T \tilde{Q}_{AA}  (\vec{\pi}_A - \vec{\pi}'_A) \right) \nn\\
&\times& \exp\left(-\frac{1}{8} 
 (\vec{\pi}_A + \vec{\pi}'_A)^T \left[ \tilde{D}^{-1}_{AA} - \frac{\tilde{D}^{-1}_{AA} P_A \tilde{D}^{-1}_{AA}}{\mathrm{tr}(P_A \tilde{ D}^{-1}_{AA})}\right] (\vec{\pi}_A + \vec{\pi}'_A) \right)\eea
Using Eq.~(\ref{eq:NorthPole}), we can now rewrite $\rho_A$ in terms of $\vec{n}_A$, $\vec{n}'_A$,
 \bea \rho_A(\vec{n}_A, \vec{n}'_A) &\propto&  \exp\left(-\frac{\rho_s}{8} (\vec{n}_A - \vec{n}'_A)^T \tilde{Q}_{AA}  (\vec{n}_A - \vec{n}'_A) \right) \nn\\
&\times& \exp\left(-\frac{\rho_s}{8} 
 (\vec{n}_A + \vec{n}'_A)^T \left[\tilde{D}^{-1}_{AA} - \frac{\tilde{D}^{-1}_{AA} P_A \tilde{D}^{-1}_{AA}}{\mathrm{tr}(P_A \tilde{ D}^{-1}_{AA})}\right] (\vec{n}_A + \vec{n}'_A) \right) \label{eq:rhoAn}\eea 
Note that $O(N)$ invariance is now restored in Eq.~(\ref{eq:rhoAn}) so it can be used on the entire sphere (and not just in the vicinity of the North pole). 

Next, we would like to deduce the spectrum of the reduced density matrix (entanglement Hamiltonian). To do this, it is convenient to parametrize $\vec{n}_A(\vec{x})$ as,
\beq \vec{n}_A( \vec{x}) = \vec{N}_A  \sqrt{1 - \frac{\chi_\alpha(\vec{x}) \chi_\alpha(\vec{x})}{\rho_s}} + \frac{\vec{E}_\alpha {\chi_\alpha(\vec{x})}}{\sqrt{\rho_s}} \label{eq:nApar}\eeq
Where $\vec{N}_A$ and $\vec{E}_\alpha$, $\alpha = 1 \ldots N-1$, are unit vectors forming an orthonormal basis, $\vec{N}_A \cdot \vec{E}_\alpha = 0$, $ \vec{E}_\alpha \cdot \vec{E}_\beta = \delta_{\alpha \beta}$. We take the fields $\chi_\alpha$ to satisfy the constraint,
\beq \int_A d^dx d^dy \,\tilde{Q}_{AA}(\vec{x}, \vec{y}) \chi_\alpha(\vec{y}) = 0 \label{eq:chiconstr}\eeq
i.e., $v^T_0 \tilde{Q}_{AA} \chi_\alpha  = 0$. Note that with this constraint, generally $v^T_0 \chi_\alpha \neq 0$. Despite this non-orthogonality, the measure on the $\vec{N}_A$, $\chi_\alpha$ space inherited from the spherical measure on $\vec{n}_A(\vec{x})$ is, to leading order in $\chi$, given by the product of a spherical measure on $\vec{N}_A$ and a flat measure on $\chi_\alpha$. Next we expand Eq.~(\ref{eq:rhoAn}) to leading order in $\chi_\alpha$, as well as in $\vec{N} - \vec{N}'$ (and also $\vec{E}_\alpha - \vec{E}'_\alpha$), obtaining
 \bea \rho_A(\vec{n}_A, \vec{n}'_A) &\propto&  \exp\left(-\frac{I}{2} (\vec{N}_A - \vec{N}'_A)^2\right)\nn\\
&\times& \exp\left(-\frac{1}{8} (\chi_\alpha - \chi'_\alpha)^T \tilde{Q}_{AA}  (\chi_\alpha - \chi'_\alpha) \right)\nn\\
&\times& \exp\left(-\frac{1}{8} 
 (\chi_\alpha + \chi'_\alpha)^T \left[\tilde{D}^{-1}_{AA} - \frac{\tilde{D}^{-1}_{AA} P_A \tilde{D}^{-1}_{AA}}{\mathrm{tr}(P_A \tilde{ D}^{-1}_{AA})}\right] (\chi_\alpha + \chi'_\alpha) \right) \label{eq:rhoAchi}\eea 
where
\beq I =  \frac{\rho_s}{4} \int_A d^d x d^dy \, \tilde{Q}_{AA}(\vec{x}, \vec{y}) \label{eq:Idef}\eeq
We see that $\rho_A$ factors into a product of density matrices in the $\vec{N}_A$ sector and in the $\chi_\alpha$ sector. 

\subsection{Tower of states sector of the entanglement Hamiltonian.}

Let us first discuss the $\vec{N}_A$ sector, where
\beq \rho_A(\vec{N}_A, \vec{N}'_A) \propto \exp\left(-\frac{I}{2} (\vec{N}_A - \vec{N}'_A)^2\right) \label{eq:rhoN}\eeq
We begin by computing the quantity $I$ in Eq.~(\ref{eq:Idef}). Using Eq.~(\ref{eq:Qexpr}), we obtain
\beq I = - \rho_s \int_A d^d x d^dx' \, \nabla^2 \tilde{D}(\vec{x}, \vec{x}') = \rho_s \int_A d^d x d^d x' \, \frac{\d}{\d x_i} \frac{\d}{\d x'_i} \tilde{D}(\vec{x} - \vec{x}') = \rho_s
\int_{\d A} dS_i \int_{\d A} d S'_i \, \tilde{D}(\vec{x} - \vec{x}') \label{eq:Ibound}\eeq
where $\d A$ represents the boundary of region $A$. By power counting, expression (\ref{eq:Ibound}) scales as $\ell^{d-1}$, with $\ell$ - the size of region $A$. A more careful analysis reveals a multiplicative logarithmic correction, whose coefficient can be extracted by focusing on the UV divergence occuring in the region $\vec{x} \to \vec{x}'$ of the integrand. The short distance behavior of $\tilde{D}(\vec{x}-\vec{x}')$ is captured by taking the infinite system limit,
\beq \tilde{D}(\vec{x}) \approx \frac{\Gamma((d-1)/2)}{4 \pi^{(d+1)/2} }\frac{1}{|\vec{x}|^{d-1}}, \quad \vec{x} \to 0 \label{eq:Dsd}\eeq 
Substituting this into Eq.~(\ref{eq:Ibound}) and zooming on the $\vec{x} \to \vec{x}'$ divergence,  
\beq I \approx  \rho_s \int_{\d A} d S \int d^{d-1} u \frac{\Gamma((d-1)/2)}{4 \pi^{(d+1)/2} }\frac{1}{|\vec{u}|^{d-1}} \approx \frac{\rho_s}{2 \pi} {\cal A} \log(\ell/a) \label{eq:Iapprox}\eeq
with ${\cal A}$ - the area (length) of the boundary of $A$. Here we have cut-off the $\vec{u}$ integral by the short-distance cut-off $a$ in the UV and by the size $\ell$ of region $A$ in the IR. Thus, $I$ scales as $I \sim \rho_s \ell^{d-1} \log(\ell/ a)$; in particular, $I \gg 1$. Now, the reduced density matrix (\ref{eq:rhoN}) is negligibly small for $(\vec{N}_A - \vec{N}'_A)^2 \gg I^{-1}$. This justifies the expansion in $\vec{N}_A - \vec{N}'_A$ we performed when computing $\rho_A$. Furthermore, using the result of the heat kernel expansion on the $(N-1)$ - dimensional unit sphere, Eq.~(\ref{HeatKer}),
we may write,
\beq \rho_A(\vec{N}_A, \vec{N}'_A) \propto \langle \vec{N}_A|\exp(-H^{\vec{N}}_E)|\vec{N}'_A\rangle \label{eq:rhoHEN}\eeq
where the entanglement Hamiltonian $H^{\vec{N}}_E$ is given by,
\beq H^{\vec{N}}_E = \frac{\vec{L}^2_A}{2 I} \label{eq:HEN}\eeq
The angular momentum operators $L_{A, a b}$ acting on the $\vec{N}_A$ space implement $O(N)$ rotations of the $\vec{N}_A$ vector and hence of $\vec{n}_{A}(\vec{x})$. Therefore, we may identify $L_{A, a b}$ with the total $O(N)$ angular momentum (spin) of the region $A$. Thus, the entanglement Hamiltonian in the $\vec{N}$ sector has the same tower of states form as the branch of the physical Hamiltonian (\ref{eq:Htowern0}). The gap in the entanglement spectrum of $H^{\vec{N}}_E$ scales as $I^{-1} \sim (\rho_s \ell^{d-1} \log \ell)^{-1}$.   

There is an instructive consistency check that we may perform at this point. The reduced density matrix $\rho_A$ should reproduce all correlation functions of operators in subsystem $A$, in particular, the fluctuations of the total angular momentum of $A$,
\beq \langle \vec{L}^2_A \rangle = \mathrm{tr}(\vec{L}^2_A \rho_A) \label{eq:fluctcheck} \eeq
Since global $O(N)$ rotations act only on the $\vec{N}_A$ sector and not on the $\chi_\alpha$ fields of Eq.~(\ref{eq:nApar}), it is sufficient to use the reduced density matrix in the $\vec{N}_A$ sector to compute the right-hand-side of Eq.~(\ref{eq:fluctcheck}).  Thus, 
\beq \mathrm{tr}(\vec{L}^2_A \rho_A) = \frac{\mathrm{tr}(\vec{L}^2_A e^{-\vec{L}^2_A/(2 I)})}{\mathrm{tr}\,e^{-\vec{L}^2_A/(2 I)}} = (N-1) I \label{eq:Lflucttr}\eeq
where in the last step we've used Eq.~(\ref{HeatKer}) together with the fact $I \gg 1$. Thus, the fluctuations of the subsystem $O(N)$ spin  scale as $\langle \vec{L}^2_A \rangle \sim \rho_s \ell^{d-1} \log \ell$, as previously obtained in Ref.~\onlinecite{Karyn}.

 At the same time, we can compute the left hand side of Eq.~(\ref{eq:fluctcheck}) by recalling that the conserved currents associated with the $O(N)$ symmetry of the non-linear $\sigma$-model (\ref{eq:Nlsmwf}) are given by,
\beq j^{\mu}_{ab} (x) = -i \rho_s \left(n_a \d_{\mu} n_b - n_b \d_{\mu} n_a\right) \label{eq:jdef}\eeq
The angular momentum of region $A$ is obtained by integrating the temporal component of the current (\ref{eq:jdef}) over region $A$,
\beq L_{A, ab} =  \int_A d^d x \, j^\tau_{ab} (\vec{x}) \eeq 
so
\beq \langle \vec{L}^2_A\rangle = \sum_{a < b} \int_A d^d x d^d y \, \langle j^\tau_{ab}(\vec{x},\tau) j^\tau_{ab}(\vec{y},\tau)\rangle\label{eq:Lfluctcorr}\eeq
To compute the correlator $\langle j^\tau_{ab}(\vec{x},\tau) j^\tau_{ab}(\vec{y},\tau)\rangle$, we use the representation  (\ref{eq:ntildepi}). Expanding $\vec{n}$ in $\tilde{\pi}_\alpha$ and also recalling that $\d_\tau \vec{n}_0 \sim L^{-d} \ll \d_\tau \tilde{\pi} \sim L^{-(d+1)/2}$, we obtain
\beq j^\tau_{ab} \approx -i \sqrt{\rho_s} (n^a_0 e^b_{\alpha} - n^b_0 e^a_{\alpha}) \d_\tau \tilde{\pi}_\alpha \eeq
so
\beq \langle j^\tau_{ab}(\vec{x}, \tau) j^\tau_{cd}(\vec{y}, \tau) \rangle = -\rho_s \langle (n^a_0 e^b_\alpha - n^b_0 e^a_\alpha)(\tau) (n^c_0 e^d_\beta - n^d_0 e^c_\beta)(\tau)\rangle \langle \d_\tau \tilde{\pi}_\alpha(\vec{x},\tau) \d_\tau \tilde{\pi}_\beta(\vec{y}, \tau) \rangle \label{eq:jjexp} \eeq
Now,
\beq \langle \d_\tau \tilde{\pi}_\alpha(\vec{x},\tau_x) \d_\tau \tilde{\pi}_\beta(\vec{y}, \tau_y)\rangle = - \delta_{\alpha \beta} \d^2_\tau \tilde{D}(\vec{x} -\vec{y}, \tau_x -\tau_y) = \delta_{\alpha \beta}(\delta^d(\vec{x}-\vec{y}) \delta(\tau_x - \tau_y) + \nabla^2 \tilde{D}(\vec{x}-\vec{y}, \tau_x - \tau_y)) \label{eq:picorr}\eeq
with $\tilde{D}(\vec{x}, \tau) = \frac{1}{V} \sum_{\vec{k} \neq 0} \int \frac{d \omega}{2\pi}  \frac{1}{\omega^2 + \vec{k}^2} e^{-i \omega \tau + i \vec{k} \cdot \vec{x}}$. In computing the equal time-correlator (\ref{eq:jjexp}) we may set $\tau_x = \tau_y + \epsilon$, $\epsilon \to 0$, to get rid of the $\delta(\tau_x - \tau_y)$ contact term in Eq.~(\ref{eq:picorr}). Then using $e^a_\alpha(\tau) e^b_\alpha(\tau) = \delta^{ab} - n^a(\tau) n^b(\tau)$, and $\langle n^a_0(\tau) n^b_0(\tau)\rangle = \frac{\delta^{ab}}{N}$,
\beq  \langle j^\tau_{ab}(\vec{x}, \tau) j^\tau_{cd}(\vec{y}, \tau) \rangle  =- \frac{2 \rho_s}{N} (\delta^{ac} \delta^{bd} - \delta^{ad} \delta^{bc}) \nabla^2 \tilde{D}(\vec{x},\vec{y})\eeq
Substituting this into Eq.~(\ref{eq:Lfluctcorr}), we obtain $\langle \vec{L}^2_A \rangle = (N-1)  I$, in agreement with Eq.~(\ref{eq:Lflucttr}). Thus, the reduced density matrix (\ref{eq:rhoHEN}) correctly reproduces the fluctuations of the $O(N)$ spin of subsystem $A$. 

\subsection{``Spin wave" sector of the entanglement spectrum.}

\label{sec:swent}

 We now discuss the $\chi_\alpha$ sector of the reduced density matrix (\ref{eq:rhoAchi}). It is convenient to make a change of variables from the $\chi_\alpha$ fields satisfying condition (\ref{eq:chiconstr}) to $\eta_\alpha$ fields satisfying,
\beq \int_A d^d x \, \eta_\alpha(\vec{x}) = 0\eeq
via
\beq \chi_\alpha = \left(1 - \frac{P_A \tilde{Q}_{AA}}{\mathrm{tr}(P_A \tilde{Q}_{AA})}\right) \eta_\alpha\eeq
Equivalently, $\eta_\alpha = (1-P_A) \chi_\alpha$. Substituting this into Eq.~(\ref{eq:rhoAchi}), 
\bea \rho_A(\eta_\alpha, \eta'_\alpha) &\propto&  \exp\left(-\frac{1}{8} (\eta_\alpha - \eta'_\alpha)^T \left[\tilde{Q}_{AA} - \frac{\tilde{Q}_{AA} P_A \tilde{Q}_{AA}}{\mathrm{tr}(P_A \tilde{Q}_{AA})}\right]  (\eta_\alpha - \eta'_\alpha) \right)\nn\\
&\times& \exp\left(-\frac{1}{8} 
 (\eta_\alpha + \eta'_\alpha)^T \left[\tilde{D}^{-1}_{AA} - \frac{\tilde{D}^{-1}_{AA} P_A \tilde{D}^{-1}_{AA}}{\mathrm{tr}(P_A \tilde{ D}^{-1}_{AA})}\right] (\eta_\alpha + \eta'_\alpha) \right) \label{eq:rhoAeta}\eea 
Note that  $  \left[\tilde{Q}_{AA} - \frac{\tilde{Q}_{AA} P_A \tilde{Q}_{AA}}{\mathrm{tr}(P_A \tilde{Q}_{AA})}\right] v_0 = 0$ and $\left[\tilde{D}^{-1}_{AA} - \frac{\tilde{D}^{-1}_{AA} P_A \tilde{D}^{-1}_{AA}}{\mathrm{tr}(P_A \tilde{ D}^{-1}_{AA})}\right]  v_0 = 0$, so we can think of both operators as restricted to act in the space orthogonal to $v_0$, i.e. in the $\eta_\alpha$ space.

The reduced density matrix (\ref{eq:rhoAeta}) is quadratic in $\eta_\alpha, \eta'_\alpha$, therefore, it can be written as an exponential of a harmonic Hamiltonian. Indeed, recall that for the 1D harmonic oscillator Hamiltonian $H_{SHO} = \frac{p^2}{2} + \frac{1}{2} \omega^2 x^2$,
\beq \langle x| e^{- H_{SHO}} |x' \rangle \propto \exp\left[-\frac{\omega}{4}(\coth( \omega/2) (x-x')^2 + \tanh( \omega/2) (x + x')^2)\right]\eeq
Generalizing to the multi-component case, $H_{SHO} = \frac12 p^T M^{-1} p  + \frac12 x^T K x$, with $x$, $p$ - vectors and $M$, $K$ - symmetric matrices,
\bea \langle x| e^{-H_{SHO}}| x'\rangle &\propto& \exp\left[-\frac{1}{4}(x-x')^T M^{1/2} \Omega \coth(\Omega/{2}) M^{1/2} (x-x')\right]\nn\\
 &\times& \exp\left[-\frac14 (x+x')^T M^{1/2} \Omega \tanh (\Omega/{2}) M^{1/2} (x+x') \right] \label{eq:SHOmult} \eea
where 
\beq  \Omega = (M^{-1/2} K M^{-1/2})^{1/2} \label{eq:Omega}\eeq
 Of course, $H_{SHO}$ can be diagonalized as $H_{SHO} = \sum_{\epsilon} \epsilon\, a^{\dagger}_{\epsilon} a_{\epsilon}$, with $\epsilon$ - eigenvalues of $\Omega$. Matching Eq.~(\ref{eq:rhoAeta}) to Eq.~(\ref{eq:SHOmult}), we obtain,

\beq \rho_A(\eta_\alpha, \eta'_\alpha) = \langle \eta_\alpha|\exp(-H^\eta_E)|\eta'_\alpha\rangle\eeq
with
\beq H^\eta_E = \frac12 (p^{\eta})^T_\alpha M^{-1} p^{\eta}_\alpha + \frac12 \eta^T_\alpha K \eta_\alpha = \sum_{\epsilon, \alpha}\, \epsilon\, a^{\dagger}_{\epsilon, \alpha} a_{\epsilon, \alpha} \label{eq:HEeta}\eeq
and
\bea M^{1/2} \Omega \coth(\Omega/{2}) M^{1/2} &=& \frac12 \left[\tilde{Q}_{AA} - \frac{\tilde{Q}_{AA} P_A \tilde{Q}_{AA}}{\mathrm{tr}(P_A \tilde{Q}_{AA})}\right]\nn\\
M^{1/2} \Omega \tanh(\Omega/{2}) M^{1/2} &=& \frac12 \left[\tilde{D}^{-1}_{AA} - \frac{\tilde{D}^{-1}_{AA} P_A \tilde{D}^{-1}_{AA}}{\mathrm{tr}(P_A \tilde{ D}^{-1}_{AA})}\right] \label{eq:cothtanh}\eea
with $\Omega$ as before related to $K$ and $M$ via Eq.~(\ref{eq:Omega}), and $\epsilon$ - eigenvalues of $\Omega$. Here, $[p^{\eta}_\alpha(\vec{x}), \eta_{\beta}(\vec{y})] = - i \delta_{\alpha \beta} (\delta^d(\vec{x} - \vec{y}) - \frac{1}{V_A})$ and $\int_A d^d x \, p^{\eta}_\alpha(\vec{x}) = \int_A d^dx\, \eta_\alpha(\vec{x}) = 0$. All operators $K$, $M$, $\Omega$ are understood to act in the space orthogonal to $v_0$. From Eqs.~(\ref{eq:cothtanh}), we obtain,
\bea M^{1/2} \coth^2(\Omega/2) M^{-1/2} &=&  \left[\tilde{Q}_{AA} - \frac{\tilde{Q}_{AA} P_A \tilde{Q}_{AA}}{\mathrm{tr}(P_A \tilde{Q}_{AA})}\right] \left[\tilde{D}^{-1}_{AA} - \frac{\tilde{D}^{-1}_{AA} P_A \tilde{D}^{-1}_{AA}}{\mathrm{tr}(P_A \tilde{ D}^{-1}_{AA})}\right]^{-1} \nn\\
&=& \left[\tilde{Q}_{AA} - \frac{\tilde{Q}_{AA} P_A \tilde{Q}_{AA}}{\mathrm{tr}(P_A \tilde{Q}_{AA})}\right] (1-P_A) \tilde{D}_{AA} (1-P_A) \eea
Thus, $\coth^2(\Omega/2)$ is similar to operator 
\beq U = \left[\tilde{Q}_{AA} - \frac{\tilde{Q}_{AA} P_A \tilde{Q}_{AA}}{\mathrm{tr}(P_A \tilde{Q}_{AA})}\right] (1-P_A) \tilde{D}_{AA} (1-P_A) \label{eq:U}\eeq
 and so the two have identical eigenvalues $\lambda$, which determine the entanglement spectrum (\ref{eq:HEeta}) through $\lambda = \coth^2 \epsilon/2$. (Of course, the eigenvalue $\lambda = 0$ of $U$ corresponding to the $v_0$ eigenvector should be omitted).

We, thus, conclude that the entire entanglement Hamiltonian is the sum of Eq.~(\ref{eq:HEN}) and Eq.~(\ref{eq:HEeta}), 
\beq H_E = \frac{\vec{L}^2_A}{2 I} + \sum_{\epsilon, \alpha} \epsilon \, a^{\dagger}_{\epsilon,\alpha} a_{\epsilon,\alpha} \label{eq:HE}\eeq 

Next, we ask how do the eigenvalues $\epsilon$ (i.e. the entanglement gaps) in the quadratic part of Eq.~(\ref{eq:HE}) scale with the subsystem size $\ell$. It is useful to first answer this question in dimension $d = 1$. Here our results are only meaningful in the case $N = 2$, i.e. for a one-dimensional superfluid or XY magnet. In this case, one can re-derive all the above results for the entanglement Hamiltonian by using a representation $\vec{n}(\vec{x}) = (\cos \phi(\vec{x}), \sin \phi(\vec{x}) )$ with $\phi(\vec{x})$ - a $2 \pi$ periodic variable.  The Lagrangian of the theory is quadratic in $\phi$,
\beq L = \frac{\rho_s}{2} (\d_\mu \phi)^2 \label{eq:Lphi} \eeq
and we assume that phase slips of $\phi$, if allowed by translational symmetry, are irrelevant in the RG sense  (i.e. the system is described by a Luttinger liquid). Now, the free compact boson theory (\ref{eq:Lphi}) in $d = 1$ is a CFT. Therefore, we can use a very general result on 1d CFTs: the entanglement spectrum for an interval of length $\ell$ embedded in an infinite system is identical to the physical spectrum of the theory on an open strip of length $\ell_{strip} = \frac{1}{\pi} \log(\ell/a)$ (if the full system is a circle of length $L$, $\ell_{strip} = \frac{1}{\pi} \log(\frac{L}{\pi a} \sin(\frac{\pi \ell}{L}))$).\cite{Holzhey} To find the physical spectrum on an open strip, we need to specify boundary conditions at the ends of the strip. By $O(2)$ symmetry, these must be free (von-Neumann) boundary conditions: $\d_{\mx} \phi = 0$. Then,
\beq H_{strip} = \frac{L^2_z}{2 \rho_s \ell_{strip}} + \sum_{\mm = 1}^{\infty} \epsilon_\mm a^{\dagger}_\mm a_\mm \label{eq:Hstrip}\eeq
with $\epsilon_\mm = \frac{\pi \mm}{\ell_{strip}}$, and $L_z \in \mathbb{Z}$ - the $O(2)$ angular momentum (spin) of the strip. We see that the $L^2_z$ part of the strip Hamiltonian (\ref{eq:Hstrip}) agrees exactly with our result for $H_E$ (\ref{eq:HE}). Indeed, the constant $I$, Eq.~(\ref{eq:Ibound}), in the 1d geometry studied is 
\beq I = 2 \rho_s (\tilde{D}(0) - \tilde{D}(\ell)) = \frac{\rho_s}{\pi} \log\left(\frac{L}{\pi a} \sin\left(\frac{\pi \ell}{L}\right)\right) = \rho_s \ell_{strip}\eeq
where we've used the expression for the 1d propagator $\tilde{D}(\mx) = -\frac{1}{2\pi} \log(2 \sin(\pi x/L))$. As for the harmonic part of the entanglement Hamiltonian, it should be possible to explicitly confirm the form $\epsilon_m = \frac{\pi m}{\ell_{strip}}$ by finding the eigenvalues of operator $U$ (\ref{eq:U}), although we will not do this here. 
An important observation is that in $d = 1$, the entanglement gaps in the $L_z$ sector and in the harmonic sector scale identically with the subsystem size as $(\log \ell/a)^{-1}$. As we will now see, this is no longer true in $d > 1$.

Proceeding to the case $d > 1$, it is convenient to consider a particular geometry, where the spatial torus is divided into two cylinders $A$ and $B$ by straight cuts at $\mathrm{x}  = 0$ and $\mathrm{x} = \ell$. We label the directions parallel to the cuts as $\mathrm{x}_{\prl}$. The cuts preserve translational invariance of the system along $\mathrm{x}_\prl$ directions. Therefore, the operators $\tilde{Q}_{AA}$, $\tilde{D}_{AA}$ (and consequently $U$) break up into sectors with definite momentum $\vec{k}_\parallel$ along the boundary of $A$ and,
\beq H^{\eta}_E = \sum_{\vec{k}_\parallel,  \mm, \alpha} \epsilon_\mm(\vec{k}_\prl) a^{\dagger}_{\vec{k}_\prl, \mm, \alpha} a_{\vec{k}_\prl, \mm, \alpha} \label{eq:HEk}\eeq
In the $\vec{k}_\parallel = 0$ sector, $U$ reduces to its form in $d = 1$. Therefore, from previous discussion, $\epsilon_\mm(\vec{k}_\parallel = 0) = \pi \mm/l_{strip}$, with $\mm \in \mathbb{N}$. In the $\vec{k}_\parallel \neq 0$ sectors we have,
\beq \tilde{D}_{AA}(\mathrm{x}, \mathrm{x}'; \vec{k}_\parallel) = \frac{1}{L_{\mx}} \sum_{k_{\mx}} \frac{1}{2 \sqrt{k^2_{\mx} + \vec{k}^2_\prl}} e^{i k_{\mx}  (\mx - \mx')}, \quad  \tilde{Q}_{AA}(\mathrm{x}, \mathrm{x}'; \vec{k}_\parallel) = \frac{1}{L_{\mx}} \sum_{k_{\mx}} {2 \sqrt{k^2_{\mx} + \vec{k}^2_\prl}} e^{i k_{\mx}  (\mx - \mx')}\eeq
and $U(\vec{k}_\parallel) = \tilde{Q}_{AA}(\vec{k}_\prl) \tilde{D}_{AA}(\vec{k}_\prl)$. Thus, $\tilde{D}_{AA}(\mathrm{x}, \mathrm{x}'; \vec{k}_\prl)$ is just the restriction of the propagator $D_{d=1}(\mx, \mx';|\vec{k}_\parallel|)$ for a one-dimensional free boson of mass $|\vec{k}_\prl|$ to region $A$. Similarly, $\tilde{Q}_{AA}$ is the restriction of $Q_{d = 1}(\mx, \mx';|\vec{k}_\parallel|) = D^{-1}_{d=1} (\mx, \mx';|\vec{k}_\parallel|)$ to region $A$. It is well known that for a free massive boson (in any dimension), the entanglement Hamiltonian is of quadratic form, $H_E = \sum_\epsilon \, \epsilon\, a^{\dagger}_\epsilon a_\epsilon$, with $\coth^2 \epsilon/2$ - eigenvalues of $Q_{AA} D_{AA}$.\cite{Peschel}  Therefore, the entanglement spectrum in the $\vec{k}_\parallel \neq 0$ sector is the same as for a one-dimensional free boson of mass $|\vec{k}_\parallel|$. While we are not aware of a general analytical form for the entangement spectrum of a 1d massive free boson for arbitrary $m \ell$, in the regime $m \ell \gg 1$, the eigenvalues $\epsilon$ are known to scale as $\epsilon \sim (\log |m a|)^{-1}$.\cite{Peschel}  Thus, for $|\vec{k}_\prl| \ell \gg 1$, $\epsilon(\vec{k}_\prl)$ in Eq.~(\ref{eq:HEk}) scales as $\epsilon(\vec{k}_\prl) \sim (\log ||\vec{k}_\prl| a|)^{-1} \sim (\log L_\parallel/a)^{-1}$, where $L_\parallel$ is the length of the $\mx_\parallel$ direction(s). Thus, we conclude that in the present geometry the entanglement gap in the quadratic sector is of order $(\log \ell/a)^{-1}$, with $\ell$ - the characteristic size of region $A$. We expect that this scaling also holds for an arbitrary geometry. As we have already discussed, the entanglement gap in the ``tower of states" sector (\ref{eq:HEN}) scales as $I^{-1} \sim (\rho_s \ell^{d-1} \log \ell/a)^{-1}$.  Thus, in $d > 1$, the entanglement gap in the ``tower of states" sector is parametrically smaller than in the harmonic sector.

Also, before we conclude this section, we note that when we take the total system size $L$ to infinity, the operator $\tilde{D}(\vec{x}-\vec{x}')$ possess a finite limit, Eq.~(\ref{eq:Dsd}) in $d > 1$,
as does $\tilde{Q}(\vec{x}-\vec{x}') = -4 \nabla^2 \tilde{D}(\vec{x}-\vec{x}')$. Therefore, the entanglement spectrum (and entanglement entropy) also have a finite limit when the subsystem size $\ell$ is kept fixed and $L \to \infty$. 

\subsection{Entanglement entropy.}
\label{sec:eewf}

Having determined the full entanglement spectrum, we are ready to compute the entanglement (Renyi) entropy. From Eq.~(\ref{eq:HE}), we obtain the $n$-the Renyi entropy,
\bea S_{n} &=& -\frac{1}{n-1} \log \mathrm{tr} (\rho^n_A)= \frac{N-1}{2} \left(\log\left(\frac{I}{2\pi}\right) +  \frac{\log n}{n -1}\right) + \log |S^{N-1}| \nn\\
&+& (N-1) \sum_\epsilon \left[\frac{1}{n-1}\left(\log \sinh (n \epsilon/2) - n \log \sinh(\epsilon/2)\right) - \log 2\right] \label{eq:Snwf} \eea
Observe that the only dependence of $S_n$ on $\rho_s$ comes from the $\frac{(N-1)}{2} \log I$ term in Eq.~(\ref{eq:Snwf}). Therefore, we conclude based on dimensional analysis,
\beq S_n = S_{UV,n}(L/a) + \Delta S_n (\rho_s L^{d-1}) \label{eq:Swfform} \eeq
with 
\beq \Delta S_n = \frac{N-1}{2} \log(\rho_s L^{d-1}) + \gamma_{\mathrm{ord},n} \label{eq:DeltaSwf} \eeq
and $\gamma_{\mathrm{ord},n}$ - a geometric constant.  In the next section, we will explicitly show that $S_{UV,n}$ has the same form as in a free boson theory,
\beq S_{UV,n} = (N-1) S^{\mathrm{free}}_{UV,n}(L/a) \label{eq:UVUV}\eeq
This result is not surprising. Indeed, when one alters the UV cut-off $a$, one integrates out short-distance fluctuations of the order parameter - i.e. the high momentum Goldstone modes $\pi_\alpha(\vec{k})$. These high momentum modes know nothing about the compactness of the order parameter and the restoration of symmetry in finite volume; they are described entirely by the free theory, $L = \frac{1}{2} (\d_{\mu} \pi_\alpha)^2$. Eq.~(\ref{eq:UVUV}) follows from this observation.

As already noted in the introduction, the structure of entanglement entropy in a free boson theory is well understood. In particular, for a subsystem with a smooth boundary in 2d and a subsystem with a flat boundary in 3d,
\beq S_{UV, n} = C_n \frac{{\cal A}}{a^{d-1}} \label{eq:UVstraight}\eeq
In this case, the geometric constant $\gamma_{\mathrm{ord},n}$ in Eq.~(\ref{eq:DeltaSwf}) is fully universal. 
On the other hand, for a subsystem in 2d whose boundary has corners,
\beq S_{UV,n} =  C_n \frac{{\cal A}}{a^{d-1}}+(N-1) \sum_i b_{\mathrm{corn},n}(\varphi_i) \log L/a \label{eq:cornerCFTwf}\eeq
Here the sum is over the corners of the boundary and the coefficient $b_{\mathrm{corn},n}(\varphi_i)$ depends on the corner angle $\varphi_i$. 
For instance, for a $90^{\circ}$ degree angle one obtains, $b_{\mathrm{corn}, n = 1}(\pi/2) \approx -0.012$ for the entanglement entropy proper and $b_{\mathrm{corn}, n = 2}(\pi/2) \approx -0.0062$ for the Renyi entropy with $n = 2$.\cite{CasiniHuerta} Note that in this case, the $\log L$ divergences (\ref{eq:cornerCFTwf}) associated with the corners and the $\log L$ divergence (\ref{eq:DeltaSwf}) associated with the compactness of the order parameter add up in the full entanglement entropy (\ref{eq:Swfform}). Also note that in this case,  $\gamma_\mathrm{ord,n}$ in Eq.~(\ref{eq:DeltaSwf}) is universal only up to an additive contribution. 

In the discussion above, we have written the scaling form (\ref{eq:Swfform}) in terms of the total system size $L$, assuming that the ratio of the subsystem size to the total system size $\ell/L$ is fixed. However, as explained in section \ref{sec:swent}, the entanglement spectrum and the entanglement entropy have a finite limit $L \to \infty$, with subsystem size $\ell$ fixed. In this limit, all the scaling forms (\ref{eq:Swfform}), (\ref{eq:DeltaSwf}) become functions of $\ell$.

Before we conclude this section, we note that given a specific geometry, the Renyi entropy may be computed by regularizing the operator $U$ (\ref{eq:U}) on the lattice, finding its eigenvalues and performing the sum over $\epsilon$ in Eq.~(\ref{eq:Snwf}). By fitting to the scaling form (\ref{eq:Swfform}), (\ref{eq:DeltaSwf}), one can then extract the geometric constant $\gamma_{\mathrm{ord}}$. We have followed this procedure for a $L\times L$ \, 2d torus partitioned into two cylinders by cuts at $\mx = 0$ and $\mx = \ell$. By studying systems with $L$ up to $300$ sites, we have verified the scaling form (\ref{eq:Swfform}), (\ref{eq:DeltaSwf}), (\ref{eq:UVstraight}) and obtained $\gamma_{\mathrm{ord}}$ for several values of the ratio $\ell/L$. Fig.~\ref{fig:gamma} presents our results  for the Heisenberg antiferromagnet case, $N = 3$.


\subsection{Entanglement in a field.}
\label{sec:wffield}
In this section, we study the entanglement spectrum and entanglement entropy of the system in the presence of a small symmetry breaking field $\vec{h}$ coupled to the order parameter as,
\beq \delta S = - M \int d^d x d\tau \,\vec{h} \cdot \vec{n} \label{hn}\eeq
Here, $M$ is the expectation value of the order parameter. In the case of a Heisenberg antiferromagnet the symmetry breaking field corresponds to a staggered magnetic field.  Using the representation (\ref{eq:NorthPole}) and expanding the action to leading order in $\vec{\pi}$
\beq S = \frac{1}{2} \int d^dx d\tau \left((\d_\mu \vec{\pi})^2 + m^2 \vec{\pi}^2\right) \label{eq:freemwf}\eeq
with the mass $m^2 = h M/\rho_s$. The static propagator of $\pi_a$ is given by
\beq \langle \pi_a(\vec{x}) \pi_b(\vec{y})\rangle = \delta_{a b} D(\vec{x},\vec{y})\eeq
\beq D(\vec{x}, \vec{y}) =  \frac{1}{V} \sum_{\vec{k}} \frac{1}{2 \sqrt{\vec{k}^2 + m^2}} e^{i \vec{k} \cdot (\vec{x} - \vec{y})} \label{eq:Dm}\eeq
We will consider the regime when the field induced Goldstone mass is much smaller than the spin-wave gap $1/L$, but much larger than the tower of states gap $\Delta_{\mathrm{tower}} \sim 1/(\rho_s L^d)$. In this regime, 
the spectrum of spin-waves with finite momentum is unaffected by the symmetry breaking field. On the other hand, the zero-momentum fluctuations of the order parameter about the staggered field are now small (of order $(\rho_s m V)^{-1/2} \ll 1$) and described by a harmonic oscillator with frequency $m$. Thus, in this field range, one may ignore the compactness of the order parameter manifold and work with the free theory (\ref{eq:freemwf}). Using $m \ll 1/L$, we may  write,
\beq D(\vec{x}, \vec{y}) \approx \frac{1}{2 m V} + \tilde{D}(\vec{x}, \vec{y}), \quad\quad Q(\vec{x}, \vec{y}) = D^{-1}(\vec{x}, \vec{y}) \approx \tilde{Q}(\vec{x}, \vec{y}) \label{eq:DQapprox} \eeq
with $\tilde{D}$, $\tilde{Q}$ given by Eqs.~(\ref{eq:Dtilde}), (\ref{eq:Qexpr}). 

As we already noted, the entanglement Hamiltonian in a free theory is quadratic, 
\beq H_E = \sum_\epsilon \epsilon\, a^{\dagger}_\epsilon a_\epsilon \label{eq:HEfree}\eeq
with $\coth^2 \epsilon/2$ - eigenvalues of ${Q}_{AA} D_{AA}$, and $Q_{AA}$, $D_{AA}$ - restrictions of $Q$, $D$ to region $A$.\cite{Peschel} By a similarity transform, $\coth^2 \epsilon/2$ are also eigenvalues of the Hermitian operator $R = Q^{1/2}_{AA} D_{AA} Q^{1/2}_{AA}$. Using (\ref{eq:DQapprox}),
\beq R = \tilde{Q}^{1/2}_{AA} \left(\frac{V_A}{2 m V} P_A + \tilde{D}_{AA} \right) \tilde{Q}^{1/2}_{AA} \label{eq:R}\eeq
We can compute the spectrum of $R$ perturbatively in $m$. Indeed, $\tilde{D}(\vec{x},\vec{y})$ scales as $1/L^{d-1}$, so the condition $m \ll 1/L$ implies that the zero mode term $\frac{V_A}{2 m V} P_A$  dominates over $\tilde{D}_{AA}$ in Eq.~(\ref{eq:R}). (If the subsystem size $\ell$ is much smaller than the total system size $L$, then $\tilde{D}_{AA}$ is given by Eq.~(\ref{eq:Dsd}), i.e. $\tilde{D}_{AA}(\vec{x}, \vec{y}) \sim 1/\ell^{d-1}$, so here we impose a slightly more stringent condition, $m \ll \frac{1}{L} \left(\frac{\ell}{L}\right)^{d-1}$). Thus, to leading order in $m$ ($O(m^{-1})$), the spectrum of $R$ is identical to that of 
\beq R_0 =\left(\frac{V_A}{2 m V} \right) \tilde{Q}^{1/2}_{AA}  P_A \tilde{Q}^{1/2}_{AA} \eeq
We see that $R_0$ is just the projector onto $\tilde{Q}^{1/2}_{AA} v_0$. Thus, $\tilde{Q}^{1/2}_{AA} v_0$ is an eigenvector of $R_0$ with eigenvalue
\beq \lambda_0 = \frac{V_A}{2 m V} v^T_0 \tilde{Q}_{AA} v_0 = \frac{1}{2 m V} \int_A  d^d x d^d y\, \tilde{Q}_{AA}(\vec{x}, \vec{y})\eeq
Repeating the calculation in Eqs.~(\ref{eq:Ibound}), (\ref{eq:Iapprox}), we have $\lambda_0 \sim \frac{1}{m V} \ell^{d-1} \log \ell \gg 1$. Hence, $\epsilon_0$ in the entanglement Hamiltonian (\ref{eq:HEfree}) corresponding to this mode, $\coth^2 \epsilon_0/2 = \lambda_0$, is approximately
\beq \epsilon_0 \approx 2 \lambda^{-1/2}_0\eeq
The other eigenvectors of $R_0$ are orthogonal to $\tilde{Q}^{1/2}_{AA} v_0$ and have eigenvalue $0$. We now use degenerate perturbation theory compute the first ($O(m^0)$) correction to these eigenvalues from the $\tilde{D}_{AA}$ term in Eq.~(\ref{eq:R}). We must diagonalize,
\beq \Delta R = (1 - P'_A) \tilde{Q}^{1/2}_{AA}  \tilde{D}_{AA} \tilde{Q}^{1/2}_{AA}  (1-P'_A)\eeq
where
\beq P'_A = \frac{\tilde{Q}^{1/2}_{AA}  P_A \tilde{Q}^{1/2}_{AA} }{\mathrm{tr}(\tilde{Q}_{AA} P_A)}\eeq
is the projector onto $\tilde{Q}^{1/2}_{AA} v_0$. Performing a similarity transformation,
\beq \tilde{Q}^{1/2}_{AA} \,\Delta R\, \tilde{Q}^{-1/2}_{AA} = \left[\tilde{Q}_{AA} - \frac{\tilde{Q}_{AA} P_A \tilde{Q}_{AA}}{\mathrm{tr}(\tilde{Q}_{AA}P_A )}\right] \tilde{D}_{AA} \left[1 - \frac{\tilde{Q}_{AA} P_A}{\mathrm{tr}(\tilde{Q}_{AA} P_A)}\right]\eeq
A further similarity transform with $S = 1 - P_A + \frac{\tilde{Q}_{AA} P_A}{\mathrm{tr}(\tilde{Q}_{AA} P_A)}$, $S^{-1} = 1 + P_A - \frac{\tilde{Q}_{AA} P_A}{\mathrm{tr}(\tilde{Q}_{AA} P_A)}$ gives,
\beq S^{-1} \tilde{Q}^{1/2}_{AA} \,\Delta R\, \tilde{Q}^{-1/2}_{AA} S = \left[\tilde{Q}_{AA} - \frac{\tilde{Q}_{AA} P_A \tilde{Q}_{AA}}{\mathrm{tr}(\tilde{Q}_{AA} P_A)}\right]  (1-P_A) \tilde{D}_{AA} (1-P_A)
\eeq
which is precisely the operator $U$ of Eq.~(\ref{eq:U}) that determines the harmonic part of the entanglement spectrum in the zero field case. Thus, we conclude that the entanglement spectra at $h = 0$, and at finite $h$ in the regime $\frac{1}{\rho_s V} \ll m \ll \frac{1}{L}$ are identical, except for the lowest branch. The Renyi entropy at finite $h$ then is
\beq S_{n} = (N-1)\left[\frac12 \log \frac{\lambda_0}{4} + \frac{\log n}{n-1}  + \sum_{\epsilon}^{} {}^{'}  \left(\frac{1}{n-1}\left(\log \sinh (n \epsilon/2) - n \log \sinh(\epsilon/2)\right) - \log 2\right)\right] \label{eq:Snmass}\eeq
where we've put the prime on the sum over $\epsilon$ to remind the reader that only eigenvalues $\coth^2 \epsilon/2$ of operator $U$ in Eq.~(\ref{eq:U}) (with the zero mode removed), should be summed over. Note that the only dependence of $S_n$ on $m$ comes from the $\log \lambda_0$ term in Eq.~(\ref{eq:Snmass}). Therefore, we conclude,
\beq S^h_n = (N-1) S^{\mathrm{free}}_{n}, \quad m \gg \frac{1}{\rho_s V}\eeq
with
\beq S^{\mathrm{free}}_{n} = S^{\mathrm{free}}_{UV,n}(L/a) + \Delta S^{\mathrm{free}}_{n}(m L) \label{eq:Sfreem}\eeq
and
\beq  \Delta S^{\mathrm{free}}_{n} = \frac{1}{2} \log\frac{1}{m L} + \gamma_{\mathrm{free},n}, \quad m \ll \frac{1}{L} \left(\frac{\ell}{L}\right)^{d-1} \label{eq:DeltaSfree}\eeq  
where $\gamma_{\mathrm{free},n}$ is a geometric constant.\footnote{For a massive free theory in $d=3$, the area law coefficient receives an additional singular contribution, resulting in a further correction to the entanglement entropy $\delta S \sim m^2 {\cal A} \log (ma)$.\cite{Fursaev} However, this correction is negligible in the regime $m L \ll 1$, and hence does not appear in our results.} Note that here and below we present all expression for a free theory with a single real scalar field. For a review of the scaling form of the UV part of the entanglement entropy $S^{\mathrm{free}}_{UV,n}(L/a)$ see section \ref{sec:eewf}. 

Now, the difference of Renyi entropy at zero field, Eq.~(\ref{eq:Snwf}), and Renyi entropy in a finite field, Eq.~(\ref{eq:Snmass}) is,
\beq S^{h = 0}_n - S^{h}_n = \frac{N-1}{2}\left[\log\left(\frac{\rho_s m V}{\pi}\right) - \frac{\log n}{n-1}\right] + \log |S^{N-1}| \label{eq:diffwf}\eeq
The difference (\ref{eq:diffwf}) is independent of the UV cut-off $a$. Therefore, we conclude that the UV parts of the Renyi entropy at zero field and at finite field are the same, as expected:
\beq S^{h = 0}_{UV,n}(L/a) = (N-1) S^{\mathrm{free}}_{UV,n}(L/a)\eeq
In the appendix, we confirm Eq.~(\ref{eq:diffwf}) using a replica-method path integral calculation in the non-linear $\sigma$-model, which serves as a futher check of the present results.

Before we conclude this section, we point out that the entanglement entropy in a free theory diverges as the mass $m \to 0$ (see Eq.~(\ref{eq:DeltaSfree})). In the physical context of a system with spontaneous symmetry breaking, the $m \to 0$ divergence is cut-off when $m$ becomes of order of the tower of states spacing $(\rho_s L^d)^{-1}$, at which point Eq.~(\ref{eq:DeltaSfree}) crosses over to the zero field result, Eq.~(\ref{eq:DeltaSwf}). Still, Eq.~(\ref{eq:DeltaSfree}) naively appears to contradict previous works on free (non-compact) theory,\cite{CasiniHuertaReview} which found that the entanglement entropy in $d > 1$ remains finite in the $m \to 0$ limit  (a weak $\log\log(1/(m\ell))$ divergence was found in $d = 1$\cite{CasiniHuerta1d}). The source of the apparent disagreement lies in the order of limits.  Our result (\ref{eq:DeltaSfree}) assumes that $m$ is taken to zero first with the total system size $L$ (and the subsystem size $\ell$) fixed. More precisely, we've assummed $m \ll \frac{1}{L} (\frac{\ell}{L})^{d-1}$. On the other hand, the order of limits in Ref.~\onlinecite{CasiniHuertaReview} is $L \to \infty$ first and then $m \to 0$, with the subsystem size $\ell$ fixed. If this order of limits is taken, then the massive propagator Eq.~(\ref{eq:Dm}) in $d > 1$ reduces to $D(\vec{x}) \to \frac{\Gamma((d-1)/2)}{4 \pi^{(d+1)/2} }\frac{1}{|\vec{x}|^{d-1}}$, becoming $m$-independent. Likewise, $Q(\vec{x}, \vec{y}) = D^{-1}(\vec{x}, \vec{y}) \to - \nabla^2 D(\vec{x}, \vec{y})$ also becomes $m$-independent. The entanglement spectrum and entanglement entropy can be deduced from eigenvalues of $Q_{AA} D_{AA}$ and remains finite in this limit: 
\beq \lim_{m \to 0} \lim_{L \to \infty} S^{\mathrm{free}}_{n} = S^{\mathrm{free}}_{UV,n}(\ell/a) + \gamma_{\mathrm{SI},n} \label{eq:gammaCFT}\eeq 
with $\gamma_{\mathrm{SI},n}$ depending only on the geometry of region $A$. In fact, this is the standard order of limits taken so the correlation functions in a free theory cluster, and the theory becomes scale invariant. For a 2d region with a smooth boundary (3d region with straight boundary), Eq.~(\ref{eq:UVstraight}) holds, and $\gamma_{\mathrm{SI}}$ is a fully universal geometric constant. Returning to the physical theory, where the order parameter is compact, we again stress the non-commuting limits:
\bea \lim_{L \to \infty} \lim_{h \to 0} S_n &=& (N-1) S^{\mathrm{free}}_{UV,n}(\ell/a) + \frac{N-1}{2} \log(\rho_s \ell^{d-1}) + \gamma_{\mathrm{ord},n}\label{eq:limLh}\\
\lim_{h \to 0} \lim_{L \to \infty} S_n &=& (N-1)  S^{\mathrm{free}}_{UV,n}(\ell/a) + (N-1) \gamma_{\mathrm{SI},n} \label{eq:limhL} \eea
Note that $Q_{AA} D_{AA}$ is distinct from the operator $U$ in Eq.~(\ref{eq:U}), therefore, the harmonic parts of entanglement spectra in the two limits (\ref{eq:limLh}) and (\ref{eq:limhL}) are different. Thus, there does not appear to be a simple relation between the constants $\gamma_{\mathrm{ord}}$ and $\gamma_{\mathrm{SI}}$.






\section{Conclusion}
\label{sec:concl}
In this paper we have demonstrated the presence of logarithmic corrections to the entanglement entropy in systems with spontaneous breaking of continuous symmetry. Such corrections have been
 recently observed in Monte-Carlo simulations.\cite{Roger2011} Our result, Eq.~(\ref{DeltaS}) gives the coefficient of the logarithmic divergence to be $b = 1$ in the case of a Heisenberg
 antiferromagnet ($N=3$) in two spatial dimensions. Presently, the Monte-Carlo simulations give $b = 0.74 \pm 0.02$.\cite{Roger2011} We believe that the difference between our exact result 
and the Monte-Carlo comes from the difficulty in extracting a subleading correction to the Renyi entropy in systems of relatively small size (up to $20\times20$ lattice sites) studied in
 Ref.~\onlinecite{Roger2011}. In addition to the coefficient of the logarithmic correction, we have also computed the constant $\gamma_{\mathrm{ord}}$ in Eq.~(\ref{DeltaS}) for the cylinder geometry studied in Refs.~\onlinecite{Roger2011, Rogertoappear}, see Fig.~\ref{fig:gamma}. The shape of $\gamma_{\mathrm{ord},n = 2}(\ell/L)$ that we obtain appears to be at least in qualitative agreement with the Monte-Carlo results of Ref.~\onlinecite{Rogertoappear}.

A curious byproduct of our work is that the Renyi entropy of a free bosonic theory diverges as the boson mass $m \to 0$, see Eq.~(\ref{eq:DeltaSfree}). This is a consequence of the shift symmetry of the free massless theory $\phi \to \phi + \mathrm{const}$.  If the field $\phi$ is non-compact, the symmetry group is $\mathbb{R}$. When one computes the expectation value of an operator $O$ which is invariant under the above symmetry, 
\beq \langle O \rangle = \frac{\int D \phi \, \psi^*(\phi) O \psi(\phi) }{\int D \phi \,\psi^* (\phi) \psi(\phi)}\eeq
the infinite factor of the group volume cancels between the numerator and denominator (here, $\psi(\phi)$ is the ground state wave-function(al), which is invariant under $\phi \to \phi + \mathrm{const}.$). However, when one computes the Renyi entropy, e.g. 
\beq S_2 = \frac{\int D \phi_A D \phi_B D \phi'_A D \phi'_B\, \psi(\phi_A, \phi_B) \psi^*(\phi'_A, \phi_B) \psi(\phi'_A, \phi'_B) \psi^*(\phi_A, \phi'_B)}{\left(\int   D \phi_A D \phi_B \,\psi^*(\phi_A, \phi_B) \psi(\phi_A, \phi_B)\right)^2}\eeq
the group volume no longer cancels - this is, essentially, the origin of the $m \to 0$ divergence (\ref{eq:DeltaSfree}). If the field $\phi$ is compact then the $m \to 0$ divergence is cut-off by the tower of states energy spacing, giving our main result (\ref{DeltaS}). Note that the divergence of the Renyi entropy in the free non-compact scalar theory (\ref{eq:DeltaSfree}) is present in any dimension, including $d = 1$. Thus, the famous result for the entanglement entropy of a free massless boson in one dimension $S = \frac{1}{3} \log L/a$ implicitly assumes compactness of the boson field. The compactification radius is secretly hidden in the short-distance cut-off $a$.
We stress that the $m \to 0$ divergence (\ref{eq:DeltaSfree}) of the free non-compact theory is present only if one takes $m$ to zero, keeping the total system size $L$ fixed. In the opposite limit, where one takes $L \to \infty$ and then $m \to 0$, keeping the subsystem size $\ell$ fixed, in $d > 1$ no divergence occurs, see Eq.~(\ref{eq:gammaCFT}). Similarly, in a physical system with spontaneous symmetry breaking, the Renyi entropy reflects the fact that the limits $L \to \infty$ and symmetry breaking field $h \to 0$ do not commmute -  see Eqs.~(\ref{eq:limLh}), (\ref{eq:limhL}).

In addition to the Renyi entropy, in the present paper we have also studied the entanglement spectrum of systems with spontaneous breaking of continuous symmetry. We have shown that the ``low-energy" part of the entanglement spectrum takes the same ``tower of states" form as the physical spectrum of the system. The level spacing of this universal part of the entanglement spectrum scales with the system size as $c/(\rho_s L^{d-1} \log L/a)$. These findings are in agreement with recent DMRG studies in one and two spatial dimensions.\cite{EntSpectCFT, EntSpectHubbard}

In this paper, we have focused on systems exhibiting spontaneous breaking of $O(N)$ symmetry down to $O(N-1)$ subgroup. It would be interesting to extend our results to other continuous symmetries and patterns of symmetry breaking. We expect that the logarithmic correction to the entanglement entropy in Eq.~(\ref{DeltaS}) is generically present with the coefficient $b = N_G/2$, where $N_G$ is the number of Goldstone modes (assuming  that all the Goldstone modes have a linear dispersion). One may also wonder what happens to the entanglement spectrum in the general case. A particularly interesting example is that of non-collinear magnetic order, where the spin rotation symmetry $SO(3)$ is fully broken. Such a non-collinear magnetic state is known to be realized by the Heisenberg model on the triangular lattice and by the $J_1- J_2$ model on the Kagome lattice with ferromagnetic next-to-nearest neighbour exchange, $J_2 < 0$. In this case, the tower-of-states spectrum of the physical Hamiltonian is that of a ``quantum top" and has a slightly more complex structure than in the case of collinear magnetic order discussed in the present paper. Remarkably, recent DMRG studies find that in this case the low-lying part of the entanglement spectrum again mirrors the physical tower of states spectrum.\cite{entNCL} We leave an analytical confirmation of this result to future work. 


\acknowledgements{We would like to thank Hong-Chen Jiang, Hyejin Ju, Matthew Hastings, Bohdan Kulchytskyy, Andreas Laeuchli, Roger Melko, Xiao-Liang Qi, Rajiv Singh and Ashvin Vishwanath for helpful discussions. We are particularly grateful to Xiao-Liang Qi for a very illuminating discussion that has stimulated our study of the entanglement spectrum. We thank the authors of Ref.~\onlinecite{Rogertoappear} for kindly sharing with us the results of their work prior to publication. The present research was initiated during the workshops ``Holographic Duality and Condensed Matter Physics" and ``Topological Insulators and Superconductors" at the Kavli Institute for Theoretical Physics, supported by the National Science Foundation under Grant No. NSF PHY05-51164. We are grateful to the workshop organizers and institute staff for their hospitality. This research was supported in part by the National Science Foundation under Grant No. NSF PHY11-25915. T. Grover is supported by a Moore foundation fellowship under the EPiQS initiative.}

\appendix

\section{Replica method.}
\label{sec:replica}
In this appendix, we calculate the entanglement entropy in the $O(N)$ non-linear $\sigma$-model using the replica  method, finding complete agreement with the wave-function method of section \ref{sec:wf}. We again start with the action (\ref{eq:Nlsmwf}).
We choose the system geometry to be a $d$-dimensional torus ${\cal T}^d$. For simplicity, we take all the directions of the torus to have length $L$. We choose the subsystem $A$ to be the cylindrical region $\ell < \mathrm{x} < L$, with $\mathrm{x}$ - one of the directions of the torus. 
 Unless otherwise noted, we assume the ratio $\ell/L$ to be finite.

As is well known, the Renyi entropy $S_n$ is given by
\beq S_n = - \frac{1}{n-1} \log \mathrm{tr} \rho^n_A = - \frac{1}{n-1} \log \frac{Z_n}{Z^n_1}\eeq
where $Z_n$ is the partition function of the system on an $n$-sheeted Riemann surface.\cite{Holzhey,Cardy} For the geometry considered here this surface is given by $(\tau, \vec{x}) \in (0, n \beta)\times{\cal T}^d$ with $\beta = 1/T$ - the inverse temperature. The following identifications need to be made on this space:
\bea (k \beta^+, \vec{x}) &\sim& ((k+1) \beta^-, \vec{x}), \quad 0 < \mathrm{x} < \ell \nn\\
(0^+, \vec{x}) &\sim& (n \beta^-, \vec{x}), \quad \ell < \mathrm{x} < L
\eea Above, $0 \leq k \leq n-1$ is an integer. 
Below, we will be interested in the low temperature limit $\beta \gg L$. 

To analyze the theory (\ref{eq:Nlsmwf}) we again use the representation (\ref{eq:NorthPole}).  
At lowest order in energy expansion, the action (\ref{eq:Nlsmwf}) then becomes a free theory of the Goldstone modes $\vec{\pi}$,
\beq S = \frac{1}{2} \int d^d x d \tau (\d_\mu \vec{\pi})^2 \label{Free}\eeq
In writing Eq.~(\ref{Free}), we have lost the information about the compact nature of the order parameter manifold. We will partially restore the compactness later in the calculation. Thus,
\beq \log Z_n = -\frac{N-1}{2} \mathrm{tr} \log (-\d^2)_n \label{eq:Znunreg}\eeq
The operator $(-\d^2)_n$ possesses a single zero mode on the $n$-sheeted Riemann surface corresponding to constant $\vec{\pi}(x)$, so it is more appropriate to write Eq.~(\ref{eq:Znunreg}) as,
\beq \log Z_n = S_0 -  \frac{N-1}{2} \mathrm{tr'} \log (-\d^2)_n \label{eq:ZnunregS} \eeq
where $S_0$ is the zero mode contribution and the prime on the trace indicates that the zero mode has been removed. We will see shortly that $S_0$ carries the information about the compactness of the order parameter. The trace in Eq.~(\ref{eq:ZnunregS}) has a UV divergence, which translates into the area-law term in the entanglement entropy. To eliminate this divergence, we will consider the difference between the entanglement entropy in zero field and in a small finite field $h$. Recall that the system in a field is described by the free non-compact massive theory (\ref{eq:freemwf}) as long as the field induced Goldstone mass $m \gg \frac{1}{\rho_s L^d}$. Thus, in this regime the difference of free energies in zero field and in a finite field is,
\beq \log Z^{h = 0}_n - \log Z^{h}_n = S_0 -  \frac{N-1}{2} \mathrm{tr'} \log (-\d^2)_n  + \frac{N-1}{2} \mathrm{tr} \log( (-\d^2)_n + m^2)\label{eq:Zreg} \eeq
Note that the contribution of eigenvalues $\lambda^2$ of $(-\d^2)_n$ with $\lambda \gg m$ cancels in Eq.~(\ref{eq:Zreg}). As in section \ref{sec:wffield}, we will take $m$ to lie in the range, $ \frac{1}{\rho_s L^d} \ll m \ll \frac{1}{L}$, so that it is sufficient to consider the contribution of eigenvalues $\lambda \ll \frac{1}{L}$. Below, we will refer to eigenvalues $0 < \lambda \ll \frac{1}{L}$ as quasi-zero modes.

Let us compute the spectrum of quasi-zero modes. We can label eigenstates of $(-\d^2)_n$ by momentum $\vec{k}_\parallel$ parallel to the boundary of subsystem $A$, which is quantized as $\vec{k}_\parallel = \frac{2\pi}{L} \vec{\mathrm{n}}_\parallel$, $\vec{\mathrm{n}}_\parallel \in \mathbb{Z}^{d-1}$. In each $\vec{k}_\parallel$ sector, $(-\d^2)_n =  (-\d^2_\perp)_{n} + \vec{k}^2_\parallel$, where the operator $(\d^2_\perp)_n$ now acts only in the two-dimensional $(\tau, \mathrm{x})$ space.  Clearly, quasi-zero modes must have $\vec{k}_\parallel = 0$. Let us now focus on the behavior of quasi-zero modes in the $(\tau, \mathrm{x})$ plane. Here, we must deal with the branch cuts at $\tau = k \beta$, $0 < \mathrm{x} < \ell$. The behavior of the eigenmodes in the vicinity of the branch cuts is expected to be nontrivial. However, for $|\tau - k \beta| \gg L$, we expect the eigenfunctions $\phi$ of $-(\d^2)_n$ to approach a linear superposition of plane wave states, $\phi \sim e^{i \omega \tau + 2 \pi i \mathrm{n}_{\mx} \mx/L}$, $\mathrm{n}_{\mx}  \in \mathbb{Z}$. For the eigenvalue $\lambda^2 = \omega^2 + (2 \pi \mathrm{n}_{\mx} /L)^2$ to be much smaller than $1/L^2$, we must choose $\mathrm{n}_{\mx} = 0$. Therefore, we expect the following asymptotic behavior of quasi-zero modes with eigenvalue $\lambda^2 = \omega^2$,
\beq \phi(\tau, \mathrm{x}) = A^+_k e^{i \omega (\tau -  k \beta)} + A^-_k e^{-i \omega (\tau -  k \beta)}, \quad \tau -  k \beta \gg L,\, (k+1) \beta - \tau \gg L \label{bc1}\eeq
Without loss of generality, we take $\omega > 0$. To study the low temperature limit $\beta \gg L$ it is convenient to cut our Riemann surface into $n$ separate sheets by defining
\beq \phi_k(\tau, \mathrm{x}) = \phi(k \beta + \tau, \mathrm{x}) \label{phikdef}\eeq
Here the variable $k$ is defined modulo $n$. Each sheet has a branch-cut at $\tau = 0$, $0 < \mx < \ell$ and the sheets are glued together along these branch cuts,
\bea \phi_k(0^+, \mx) &=& \phi_{k+1}(0^-,\mx), \quad 0 < \mx < \ell \nn\\
\d_\mu \phi_k(0^+, \mx) &=& \d_\mu \phi_{k+1}(0^-,\mx), \quad 0 < \mx < \ell \label{cut}\eea
At finite $\beta$, we should also glue the sheets via
\beq \phi_k(\tau, \mx) = \phi_{k+1}(\tau - \beta, \mx) \label{glue2}\eeq
However, since for $|\tau| \gg L$, $\phi_k$ approaches the plane wave states (\ref{bc1}), we will take the $\tau$ coordinate in each sheet to run from $-\infty$ to $\infty$ and implement Eq.~(\ref{glue2}) via a boundary condition,
\bea \phi_k(\tau, \mx) &=& A^+_k e^{i \omega \tau} + A^-_k e^{-i \omega \tau}, \quad \tau \to \infty\nn\\
\phi_k(\tau, \mx) &=& A^+_{k-1} e^{i \omega (\tau + \beta)} + A^-_{k-1} e^{-i \omega (\tau + \beta)}, \quad \tau \to -\infty\label{bc2}\eea
The corrections to this approximation are expected to be of order $e^{-\beta L}$. Thus, we must solve
\beq -\d^2 \phi_k = \omega^2 \phi_k \label{Helmholz}\eeq
subject to the boundary conditions (\ref{cut}), (\ref{bc2}). 

We may further simplify Eq.~(\ref{Helmholz}) in the quasi-zero mode limit $\omega \ll 1/L$. Indeed, in the vicinity of the branch points, we expect $\phi$ to vary on the length-scale $L$. Thus, the typical contributions to the left hand side of Eq.~(\ref{Helmholz}) are of order $1/L^2$ and we may set the right hand side of Eq.~(\ref{Helmholz}) to zero,
\beq -\d^2 \phi_k = 0 \label{Laplace} \eeq
We expect the approximation (\ref{Laplace}) to hold as long a $|\tau| \ll 1/\omega$. On the other hand, once $|\tau| \gg L$ the asymptotic forms (\ref{bc2}) become valid, so we should solve Eq.~(\ref{Laplace}) subject to the boundary conditions,
\bea \phi_k(\tau, \mx) &=& (A^+_k + A^-_k) +  i (A^+_k - A^-_k) \omega \tau, \quad \tau \to \infty\nn\\
\phi_k(\tau, \mx) &=& A^+_{k-1} e^{i \omega \beta} + A^-_{k-1} e^{-i \omega \beta} + i (A^+_{k-1} e^{i \omega \beta} - A^-_{k-1} e^{-i \omega \beta}) \omega \tau, \quad \tau \to -\infty\label{bc3}\eea
together with the gluing condition (\ref{cut}). 

We can solve the Laplace equation (\ref{Laplace}) using conformal mapping. Let us define $z = \mx + i \tau$ on each sheet of the Riemann surface. Topologically, each sheet is a cylinder due to the periodicity of $\mx$. We begin by mapping each cylinder to a complex plane using $w = e^{2 \pi i z/L}$. This maps the branch cut at $\tau = 0$, $0 < \mx < \ell$ to an arc of a unit circle with angle $0 < \theta < 2 \pi \ell/L$. Moreover, $\tau = \infty$ is mapped to the origin and $\tau = - \infty$ to the point at infinity. Next, we apply the following map to each sheet, 
\beq \zeta = e^{-\pi i \ell/L} \frac{w-e^{2 \pi i \ell/L}}{1-w} \eeq
This maps the branch cut into the positive $x$-axis. Moreover, $\tau = \pm \infty$ is now mapped to $-e^{\pm \pi i \ell/L}$. Finally, we map the $n$-sheets into a single sheet using $s = \zeta^{1/n}$. More precisely, for $\zeta$ on the $k$-th sheet with $\zeta = r e^{i \theta}$, $r > 0$, $0 < \theta < 2\pi$, we define $\zeta^{1/n} = r^{1/n}e^{i (\theta + 2 \pi k)/n}$. With this definition and the gluing (\ref{cut}), there are no branch cuts in the $s$ plane. Note that the points $\tau = \pm \infty$ on the $k$th sheet now map into $s^{\pm}_k = e^{\pi i(2 k+1 \pm \ell/L)/n}$. The boundary conditions (\ref{bc3}) imply logarithmic divergences for $s \to s^{\pm}_k$. Note that these are the only singularities that can occur in the $s$-plane, including the point $s = \infty$. Indeed, $s = \infty$ is the image of the branch point $\tau = 0$, $\mx = 0$ at which we expect no singularity. Therefore,
\beq \phi(s) = \sum_{k} \left(c^+_k \log|s-s^+_k| + c^-_k \log|s-s^-_k|\right) + C, \label{phis}\eeq
Note that the absence of a singularity at $s = \infty$ implies,
\beq \sum_{k} \left(c^+_k + c^-_k\right) = 0 \label{neut1}\eeq
The solution (\ref{phis}) has the following asymptotic behavior as $\tau \to \pm \infty$ on $k$th sheet,
\bea \phi_k(\tau, \mx) &\stackrel{ \tau \to \infty }{\to}& - \frac{2 \pi \tau}{L}  c^+_k + \log\left(\frac{2}{n} \sin(\pi \ell/L)\right) c^+_k + \sum_{l\neq k} c^+_l \log |s^+_k - s^+_l| + \sum_l c^-_l \log|s^+_k -s^-_l| + C,\nn\\
\phi_k(\tau, \mx) &\stackrel{ \tau \to -\infty }{\to}&  \frac{2 \pi \tau}{L}  c^-_k + \log\left(\frac{2}{n} \sin(\pi \ell/L)\right) c^-_k + \sum_{l} c^+_l \log |s^-_k - s^+_l| + \sum_{l\neq k} c^-_l \log|s^-_k -s^-_l| + C\nn\\
\label{phikass}\eea
Imposing the boundary conditions (\ref{bc3}) we obtain,
\bea c^+_k &=& \frac{-i \omega L}{2\pi} (A^+_k - A^-_k) \\
c^-_k &=& \frac{i \omega L}{2\pi}(A^+_{k-1} e^{i \omega \beta} - A^-_{k-1} e^{-i \omega \beta}) \label{cA}\eea
and
\bea A^+_k  + A^-_k - C &=& \log\left(\frac{2}{n} \sin(\pi \ell/L)\right) c^+_k + \sum_{l\neq k} c^+_l \log |s^+_k - s^+_l| + \sum_l c^-_l \log|s^+_k -s^-_l|\nn\\
A^+_{k-1} e^{i \omega \beta} + A^-_{k-1} e^{-i \omega \beta}  - C &=& \log\left(\frac{2}{n} \sin(\pi \ell/L)\right) c^-_k + \sum_{l} c^+_l \log |s^-_k - s^+_l| + \sum_{l\neq k} c^-_l \log|s^-_k -s^-_l|\nn\\\label{constbc} \eea
Note that from (\ref{cA}) the right-hand side of Eq.~(\ref{constbc}) is suppressed compared to the left-hand side by a factor of $\omega L$, and at leading order may be dropped. Thus, we must solve
\bea && A^+_k  + A^-_k  = C \label{cond1}\\
&& A^+_{k-1} e^{i \omega \beta} + A^-_{k-1} e^{-i \omega \beta} = C\label{cond2}\\
&& (1- e^{i \omega \beta}) \sum_k A^+_k - (1- e^{-i \omega \beta}) \sum_k A^-_k = 0 \label{neut2}\eea
Here, Eq.~(\ref{neut2}) comes from the condition (\ref{neut1}). To further simplify the above equations, we observe that our original problem is symmetric under the time translation $\tau \to \tau + \beta$, which generates a cyclic permutation of the $n$ sheets. In the $s$ plane it maps $s \to e^{2 \pi i/n}s$. We choose $\phi$ to be an eigenstate of this symmetry  with eigenvalue $e^{2 \pi i p/n}$, which implies,
\beq A^{\pm}_k = A^{\pm} e^{2 \pi i p k/n} \eeq 
For $p \neq 0$, the symmetry also constrains $C = 0$. Moreover, in this case Eq.~(\ref{neut2}) is trivially satisfied. Then, solving Eqs.~(\ref{cond1}),(\ref{cond2}) we obtain $A_+ = - A_-$ and the condition $\sin (\omega \beta) = 0$, {\it i.e.} 
\beq \omega = \frac{\pi j}{\beta},\, j \in \mathbb{N},\, \quad p \neq 0 \label{pne0}\eeq
Thus, for ``momentum'' $p \neq 0$, one mode is present for each $j$ in Eq.~(\ref{pne0}). On the other hand, for $p = 0$, we obtain $C = A_+ + A_-$ and $e^{i \omega \beta} = 1$, {\it i.e.},
\beq \omega = \frac{2 \pi j}{\beta},\, j \in \mathbb{N},\,\quad p = 0\label{p0}\eeq
In this case, $A_+$ and $A_-$ are independent, so two  modes are present for each $j$ in Eq.~(\ref{p0}). The explicit form of the corresponding eigenstates is,
\beq \phi = A^+ + A^- - \frac{i \omega L}{2 \pi} (A^+ - A^-) \sum_k \log\left|\frac{s-s^+_k}{s-s^-_k}\right| = A^+ + A^- + i \omega \tau (A^+ - A^-)\label{p0tau}\eeq
We have inverted the conformal mapping in the last step. Upon continuing Eq.~(\ref{p0tau}) from $|\omega \tau| \ll 1$ to the full range of $\tau$ we obtain,
\beq \phi = A^+ e^{i \omega \tau} + A^- e^{-i \omega \tau}\label{p0full}\eeq
For $\omega$ satisfying (\ref{p0}), Eq.~(\ref{p0full}) is actually the exact eigenstate of $-\d^2$ on the $n$-sheeted surface. In contrast, the solutions with $p \neq 0$ that we've constructed are not exact. The leading correction to these eigenstates can be calculated by keeping the terms on the right-hand side of Eqs.~(\ref{constbc}). This gives an $O(L/\beta)$ corrections to the eigenvalues (\ref{pne0}). We have verified that these corrections do not modify any of the leading order results below for the Renyi entropy. 

We are now ready to compute the difference of partition functions (\ref{eq:Zreg}).
We begin with the zero mode contribution $S_0$. Physically, the presence of the zero-mode indicates the degeneracy of the vacuum manifold. Indeed, in Eq.~(\ref{eq:NorthPole}) we have expanded our field $\vec{n}$ around a fixed direction. However, at $h = 0$, we should integrate over all directions of $\vec{n}$. Thus, $S_0$ should be converted into an integral over the order parameter manifold. Locally the manifold can be parameterized by
\beq \vec{\pi}(x) = \vec{\pi}_0\eeq
and 
\beq S_0 = \int d^{N-1}\vec{\pi}_0 \sqrt{\mathrm{det} g} \eeq
with the metric
\beq g_{ab} = \frac{1}{2\pi}\int d^d x d \tau \frac{\d \vec{\pi}(x)}{\d \pi^a_0}\cdot \frac{\d \vec{\pi}(x)}{\d \pi^b_0} = \frac{n \beta L^d}{2\pi} \delta_{ab} \eeq
Thus,
\beq S_0 = \left(\frac{n \beta L^d}{2\pi}\right)^{(N-1)/2} \int d^{N-1} \vec{\pi}_0 \approx  \left(\frac{n \beta \rho_s L^d}{2\pi}\right)^{(N-1)/2} \int d \vec{n} = \left(\frac{n \beta \rho_s L^d}{2\pi}\right)^{(N-1)/2} |S^{N-1}| \label{S0res}\eeq 
where we have gone from a local integral over $\vec{\pi}_0$ to a global integral over the order parameter orientation $\vec{n}$. 

Next, consider the contribution of quasi-zero modes,
\bea \log Z^{qz, h =0}_n - \log Z^{qz, h}_n &=& - \frac{N-1}{2} (\mathrm{tr}' \log (-\d^2_n) - \mathrm{tr}' \log ((-\d^2)_n + m^2)) \nn\\ &=& -\frac{N-1}{2} \Bigg[(n-1) \sum_{j=1}^{\infty} \left(\log(\pi j/\beta)^2 - \log\left( (\pi j/\beta)^2 + m^2\right)\right) \nn\\
&+& 2 \sum_{j=1}^{\infty} \left(\log(2\pi j/\beta)^2 - \log\left( (2\pi j/\beta)^2 + m^2\right)\right)\Bigg]\eea
The first sum in the square brackets gives the contribution of modes (\ref{pne0}) with $p \neq 0$, while the second sum gives the contribution of modes (\ref{p0}) with $p = 0$. Using the standard relation,
\beq -\frac{1}{2} \sum_{j=-\infty}^{\infty} \left(\log ((2\pi j/\beta)^2 + \omega^2) - \log((2\pi j/\beta)^2 + \omega'^2\right) = \log Z_{ho}(\omega, \beta) - \log Z_{ho}(\omega',\beta)\eeq
where $Z_{ho}(\omega, \beta)$ is the partition function of a harmonic oscillator with frequency $\omega$ at inverse temperature $\beta$,
\beq Z_{ho} = \frac{1}{2 \sinh (\beta \omega/2)}\eeq
we obtain in the limit $\beta m \gg 1$,
\beq \log Z^{qz, h = 0}_n- \log Z^{qz, h}_n= - \frac{N-1}{2} \left(-n \beta m + (n+1) \log (\beta m) + (n-1) \log 2\right) \label{Zqzres}\eeq
Combining (\ref{Zqzres}) with (\ref{S0res}) and taking the contribution of the zero mode to the second trace in Eq.~(\ref{eq:Zreg}) into account, we obtain
\beq \log Z^{h = 0}_n - \log Z^{h}_n = \frac{N-1}{2} \left[ n \beta m + \log(n \rho_s L^d/(2\pi\beta)) - (n-1) \log(2\beta m)\right] + \log|S^{N-1}| \label{Znres}\eeq
which gives the difference of Renyi entropies,
\bea S^{h = 0}_n - S^h_n&=& - \frac{1}{n-1} \left( \log \frac{Z^{h = 0}_n}{(Z^{h= 0}_1)^n}- \log \frac{Z^{h}_n}{(Z^{h}_1)^n}\right) \nn\\
&=& \frac{N-1}{2} \left[\log (\rho_s L^d m/\pi) - \frac{\log n}{n-1}\right] + \log|S^{N-1}| \label{DeltaSfin}\eea
Eq.~(\ref{DeltaSfin}) exactly agrees with our result (\ref{eq:diffwf}) obtained using the wave-function method. Recall that $S^h_n$ is just the Renyi entropy of a free massive theory, whose scaling form is given by Eq.~(\ref{eq:Sfreem}). Thus, from the difference (\ref{DeltaSfin}) we can obtain the scaling form for the entanglement entropy in zero field, Eq.~(\ref{DeltaS}).

Before we conclude this section, we note that our path-integral calculation above is strictly justified only for temperature much greater than the tower of states energy spacing, $T \gg \Delta_{\mathrm{tower}} \sim \frac{1}{\rho_s L^d}$. This can be seen from studying the physical free energy $F(T) = -\frac{1}{\beta} \log Z_1$. Observe that Eq.~(\ref{Znres}) gives, 
\beq (F(T) - F(0))|_{h=0}  \approx (F(T) - F(0))|_{h=0} - (F(T) - F(0))|_{h} = - T \left(\frac{N-1}{2} \log \frac{\rho_s L^d T}{2\pi} + \log|S^{N-1}|\right) \label{eq:F}\eeq
where the first approximate equality holds as we are only considering temperatures $T \ll m$. Eq.~(\ref{eq:F}) is in agreement with the free energy of the physical Hamiltonian (\ref{eq:H}) for $T \gg 1/(\rho_s L^d)$, but not for $T \ll 1/(\rho_s L^d)$, where the physical $F(T) - F(0)$ is exponentially suppressed (note that in the temperature range $T \ll 1/L$ considered here, only the tower of states sector, Eq.~(\ref{eq:Htowern0}), contributes to the free energy). The technical reason for the disagreement between the present path-integral treatment and the Hamiltonian calculation is that the former takes the compactness of the order parameter into account only in the treatment of zero modes, whereas the finite frequency modes are still described by a non-compact free boson theory. This approximation is reliable for $T \gg 1/(\rho_s L^d)$ where the fluctuations of $\vec{n}$ corresponding to the finite frequency modes are small, but breaks down for $T$ below $1/(\rho_s L^d)$ where these fluctuations are large. However, we observe that while our free energy has an unphysical temperature dependence for $T \ll 1/(\rho_s L^d)$, this temperature dependence disappears in the expression for the Renyi entropy (\ref{DeltaSfin}). Thus, one may guess that the result (\ref{DeltaSfin}) is correct not only for $1/(\rho_s L^d) \ll T \ll c/L$, but actually for all $T \ll c/L$. This is confirmed by the zero temperature wave-function calculation of section \ref{sec:wf}. Furthermore, in appendix \ref{neq2} we furthter confirm this by a replica method calculation for the special case $N = 2$, where the compactness of the order parameter manifold can be taken into account exactly for all temperatures.

\section{Replica method calculation of Renyi entropy for $N=2$ at $T = 0$} \label{neq2}

In this appendix we show that Eq.~(\ref{DeltaSfin}), indeed, remains correct at $T = 0$ by an explicit calculation in the case $N =2$. Here, we may use the angular representation
\beq \vec{n} = (\cos \phi,  \sin \phi)\eeq
with the action,
\beq S = \frac{\rho_s}{2} \int d^dx d \tau (\d_\mu \phi)^2 \label{Stheta}\eeq
The compactness of the order parameter manifold is now expressed through the identification $\phi \sim \phi + 2 \pi$. 

Since the theory (\ref{Stheta}) is free, we can simply repeat our previous calculation of the entanglement entropy. The only additional complication is that we should restore the periodicity of the variable $\phi$. This is accomplished by summing over all winding number configurations of $\phi$. More precisely, each field configuration on the $n$-sheeted Riemann surface is characterized by the winding numbers $r_k$, with
\beq \int_{k \beta}^{(k+1) \beta}  d \tau \,\d_\tau \phi(\tau, x) = 2 \pi r_k, \quad 0 < \mx < \ell\eeq
and
\beq \int_{0}^{n \beta} d\tau \,\d_\tau\phi(\tau, x) = 2 \pi \sum_k r_k, \quad \ell < \mx < L  \label{wind2}\eeq
Here we are assuming that the field $\phi(\tau, x)$ has no vortices in the $(\tau, \mx)$ plane, also known as phase-slips. Such phase-slips have an action proportional to $L^{d-1}$ and are, hence, exponentially suppressed.\footnote{The phase-slips are also often suppressed by lattice symmetries.} Ordinary spatial vortices are also exponentially suppressed as $T \to 0$. 

Any field in the sector with a given set of winding numbers may be written as
\beq \phi(x) = \phi_r(x) + \delta \phi(x) \eeq
Here, $\phi_r$ is a reference field carrying winding numbers $\{r_k\}$ and $\delta \phi(x)$ is a field with all winding numbers equal to zero. We choose $\phi_r$ to satisfy
\beq \d^2 \phi_r = 0 \label{Laplacer}\eeq
We will discuss the solution to this equation shortly. Then
\beq S[\phi] = S[\phi_r] + S[\delta \phi]\eeq 
Thus, we may perform the integral over $\delta \phi$, to obtain
\beq Z^{h = 0}_n = Z^{r = 0}_n W_n \eeq
with
\beq W_n = \sum_r e^{-S[\phi_r]}\label{Wn}\eeq
We have already computed the partition function $Z^{r=0}_n$ in the zero winding number sector - it is given by Eq.~(\ref{Znres}). It remains to compute $S[\phi_r]$.

Let us solve Eq.~(\ref{Laplacer}) in a sector with winding numbers $\{r_k\}$. We take $\phi(x)$ to be independent of the variables along the subsystem boundary and focus on the $(\mx, \tau)$ plain. We again introduce fields $\phi_k$ via Eq.~(\ref{phikdef}). As before the fields are glued along the branch-cuts at $\tau = k \beta$ using Eq.~(\ref{cut}). However, to introduce the winding numbers instead of Eq.~(\ref{glue2}) we now impose,
\beq \phi_k(\tau) = \phi_{k+1}(\tau - \beta) + 2 \pi r_k \label{bcr}\eeq
As before, we work in the limit $\beta \gg L$, such that the $\tau$ variable in each sheet runs from $(-\infty, \infty)$. The solution to the Laplace equation with the above boundary conditions is then given by Eq.~(\ref{phis}). In the present case, we may set $C = 0$ (a finite $C$ can be absorbed into $\delta\phi$). Using the expansion (\ref{phikass}) and imposing boundary conditions (\ref{cut}), (\ref{bcr}) we find,
\beq c^-_k = - c^+_{k-1} \label{cpmr}\eeq
and
\beq c^+_k = -\frac{L}{\beta} r_k + O\left(\frac{L^2}{\beta^2}\right) \label{ckrk}\eeq
Note that due to the relation (\ref{cpmr}), the constraint (\ref{neut1}) is automatically satisfied.

We are now ready to calculate the action $S[\phi_r]$ in each winding number sector. We may write,
\beq S[\phi_r] = \frac{\rho_s L^{d-1}}{2} \int_0^{n \beta} d\tau \int_0^{L} d\mx (\d_\mu \phi)^2 =  \frac{\rho_s L^{d-1}}{2} \sum_k \int_{-\beta/2}^{\beta/2} d\tau \int_0^{L} d\mx (\d_\mu \phi_k)^2 \label{Srsumk}\eeq
Here we've broken up the integral into contributions from each sheet. We now integrate by parts and use Eq.~(\ref{Laplacer}) to reduce Eq.~(\ref{Srsumk}) to an integral over the boundary of each sheet,
\bea  S[\phi_r] &=& \frac{\rho_s L^{d-1}}{2} \sum_k \int_k dS_\mu \, \phi_k \d_\mu \phi_k \nn
\\ &=& \frac{\rho_s L^{d-1}}{2} \sum_k \Bigg(\int_0^L d\mx ((\phi_k \d_\tau \phi_k)(\mx, \beta/2) - (\phi_k \d_\tau \phi_k)(\mx, -\beta/2))\label{phidphi1}\\
&& ~~~~~~~~~~~~~-\int_0^\ell ((\phi_k \d_\tau \phi_k)(\mx, 0^+) - (\phi_k \d_\tau \phi_k)(\mx, 0^-))\Bigg)\label{phidphi2}\eea
The contribution (\ref{phidphi2}) comes from the cuts at $\tau = k \beta$ and vanishes by Eq.~(\ref{cut}). On the other hand, using Eq.~(\ref{bcr}), the contribution (\ref{phidphi1}) becomes
\beq S[\phi_r] = \frac{\rho_s L^{d-1}}{2} \sum_k (2 \pi r_k) \int_0^L d\mx\, \d_\tau \phi_k(\mx, \beta/2) = \frac{\rho_s L^{d-1}}{2} \sum_k (2 \pi r_k)\left(-{2 \pi c^+_k}\right) \approx  \frac{\rho_s L^d}{2 \beta} \sum_k (2 \pi r_k)^2\eeq
where we've used Eqs.~(\ref{phikass}),(\ref{ckrk}) and dropped a correction of order $L/\beta$ in the last step. Therefore, the winding number factor $W_n$ in Eq.~(\ref{Wn}) satisfies,
\beq W_n \approx \nu^n(e^{-2\pi^2 \rho_s L^d T})\eeq
with $\nu(q) = \sum_{r=-\infty}^{\infty} q^{r^2}$ - the Jacobi theta function. Observe that
\beq \frac{W_n}{W^n_1} = 1 +O(L/\beta)\eeq
Hence, the winding number contribution does not to leading order modify our previous result for the entanglement entropy Eq.~(\ref{DeltaSfin}) for $T \ll 1/L$. On the other hand, the partition function on the $n$-sheeted Riemann surface is modified by the inclusion of $W_n$,
\bea \log Z^{h = 0}_n - \log Z^{h}_n &\stackrel{N = 2}{=}& \frac{1}{2}\bigg(n \beta m + \log(n\rho_s L^d/\beta) - (n-1) \log(2 \beta m)  \nn\\
&+& 2 n \log \nu(\exp(-2 \pi^2 \rho_s L^d T)) + \log (2 \pi)\bigg) \label{Znwind}\eea
In the limit, $T \gg \frac{1}{\rho_s L^d}$, $W_n \to 1$ and the partition function (\ref{Znwind}) reduces to our previous result (\ref{Znres}) as expected. However, using the identity $\nu(\exp(-\alpha)) = \sqrt{\pi/\alpha}\, \nu(\exp(-\pi^2/\alpha))$, we may also rewrite Eq.~(\ref{Znwind}) as,
\beq \log Z^{h = 0}_n - \log Z^{h}_n = \frac{1}{2} \left(n \beta m - (n-1) \log (4 \pi \rho_s L^d m) + \log n + 2n \log \nu(\exp(-1/(2 \rho_s L^d T)))\right)\eeq
so that for $T \ll 1/(\rho_s L^d)$,
\beq \log Z^{h = 0}_n - \log Z^{h}_n \to \frac{1}{2} \left(n \beta m - (n-1) \log (4 \pi \rho_s L^d m) + \log n \right)\eeq
Thus, we see that the singular logarithmic dependence of (\ref{Znres}) in the limit $T \to 0$ disappears once the sum over winding numbers is performed.

\end{document}